# Simultaneous use of Individual and Joint Regularization Terms in Compressive Sensing: Joint Reconstruction of Multi-Channel Multi-Contrast MRI Acquisitions

Emre Kopanoglu[1,2,*], Alper Güngör[2,3], Toygan Kilic[3,4], Emine Ulku Saritas[3,4,5], Kader K. Oguz[4,6], Tolga Çukur[3,4,5,§], H. Emre Güven[2,§]

[1] Cardiff University Brain Research Imaging Centre (CUBRIC), School of Psychology, Cardiff University, Cardiff, UK

[2] ASELSAN Research Center, Ankara, Turkey

[3] Department of Electrical and Electronics Engineering, Bilkent University, Ankara, Turkey

[4] National Magnetic Resonance Research Center (UMRAM), Bilkent University, Ankara, Turkey

[5] Neuroscience Program, Sabuncu Brain Research Center, Bilkent University, Ankara, Turkey

[6] Department of Radiology, Hacettepe University, Ankara, Turkey

**Running Head:** SIMIT: Joint Reconstruction of Multi-Channel Multi-Contrast MRI Images

*****: To whom correspondence should be addressed.

**Corresponding Author:** Emre Kopanoglu

**Address:** CUBRIC, Cardiff University, Maindy Road, Cardiff, CF24 4HQ, UK

**e-mail:** emre.kopanoglu@gmail.com

**§**: These senior authors co-supervised this study.

Words: 226 (abstract), 5969 (body)

Figures: 12

## ABSTRACT

Multi-contrast images are commonly acquired together to maximize complementary diagnostic information, albeit at the expense of longer scan times. A time-efficient strategy to acquire high-quality multi-contrast images is to accelerate individual sequences and then reconstruct undersampled data with joint regularization terms that leverage common information across contrasts. However, these terms can cause features that are unique to a subset of contrasts to leak into the other contrasts. Such leakage-of-features may appear as artificial tissues, thereby misleading diagnosis. The goal of this study is to develop a compressive sensing method for multi-channel multi-contrast magnetic resonance imaging (MRI) that optimally utilizes shared information while preventing feature leakage. Joint regularization terms group sparsity and colour total variation are used to exploit common features across images while individual sparsity and total variation are also used to prevent leakage of distinct features across contrasts. The multi-channel multi-contrast reconstruction problem is solved via a fast algorithm based on Alternating Direction Method of Multipliers. The proposed method is compared against using only individual and only joint regularization terms in reconstruction. Comparisons were performed on single-channel simulated and multi-channel *in-vivo* datasets in terms of reconstruction quality and neuroradiologist reader scores. The proposed method demonstrates rapid convergence and improved image quality for both simulated and *in-vivo* datasets. Furthermore, while reconstructions that solely use joint regularization terms are prone to leakage-of-features, the proposed method reliably avoids leakage via simultaneous use of joint and individual terms, thereby holding great promise for clinical use.

## INTRODUCTION

Multiple images of the same anatomy under the influence of different contrasts are often acquired in magnetic resonance imaging (MRI) to accumulate diagnostic information. Common examples include multi-contrast imaging with T1, T2, or PD weighting, parametric mapping, and diffusion weighted imaging. However, with each acquisition lasting several minutes, MRI exams can become impractically long and costly. Prolonged scan times also increase susceptibility to patient motion and necessitate cumbersome motion correction or image registration procedures. Therefore, scan-time reduction techniques are direly needed to limit cost, patient discomfort, and motion with increasing number of acquisitions.

Accelerated imaging approaches including parallel imaging (PI) [1-6], multi-slice imaging [7-9] localizing the excitation volume [10-17], dephasing outer volumes [18,19], localizing encoding to a sub-volume [20-22] and compressive sensing (CS) [23-45] are promising solutions. Among these, CS has gained prominence in the last decade, as it does not require complicated excitation pulses with increased specific absorption rate (SAR) or additional hardware. Furthermore, CS is compatible with alternative approaches such as parallel imaging [32,41] and simultaneous multi-slice imaging [46].

Conventional CS techniques process each acquisition in a multi-contrast protocol individually [25-28]. Yet, although tissues may appear at different signal levels in separate contrasts, the underlying tissue structure is shared among multi-contrast acquisitions. As such, multi-contrast images share common tissue boundaries, and they are likely compressible in similar transform domains. These observations have motivated researchers to investigate the benefits of jointly reconstructing multiple images of the same anatomy [29-40]. Proposed application domains include dynamic MRI reconstructions that handle single-contrast acquisitions across time [35,36], multi-coil MRI reconstructions [32], multi-echo MRI reconstructions that handle repeated acquisitions with minor changes in contrast [33,37], fat-water separation [38], multi-contrast MRI reconstructions that process multiple distinct contrasts [29-31], and even multi-modality reconstructions [34].

Joint reconstructions aim to utilize the information shared across contrasts to improve image quality. Sparse recovery during joint reconstructions has been attempted with a multitude of regularization terms in literature. A group of studies have focused on aggregation of individual regularization terms on each separate image such as the well-known $\ell_1$-sparsity [47] and Total Variation (TV) [48] terms. In Ref. [32], sparsity was promoted simultaneously across multiple receive channels by imposing $\ell_1$-sparsity on concatenated multi-channel dynamic MRI data. Ref. [38] jointly reconstructed water and fat images from a multi-echo acquisition by minimizing the sum of individual regularization functions on each image. Ref. [49] performed a quasi-joint reconstruction by spatially weighing the individual $\ell_1$-sparsity and Total Variation of an image using structural information extracted from a prior individually reconstructed image. The performance improvement with joint reconstruction depends on how shared information is leveraged against the information unique to each contrast. Classical individual regularization terms help preserve unique information in each contrast without leakage of distinct features across images, but reconstructions can be sub-optimally sensitive to shared information across contrasts. Meanwhile, joint regularization terms such as group sparsity [50] and Colour TV [51] that enforce $\ell_1$-sparsity and total variation on multiple images simultaneously have provided useful in several applications including parametric mapping [40], diffusion tensor imaging [39], multi-echo T2-weighted imaging [29,33] and multi-contrast imaging [30,31,52,53]. Ref. [35] minimized the nuclear norm to exploit the temporal correlations whereas Ref. [36] used Frobenius and nuclear norms in a blind compressed sensing approach to dynamic MRI. Variations of TV regularization such as total generalized variation [34], parallel sets [54] and weighted joint (colour) TV [55,56] have also been used for jointly reconstructing MR and PET images [34,54,56] or for multi-contrast MRI [57,58]. Joint regularization terms boost sensitivity for features that are common across acquisitions, but as a result they can reduce sensitivity for features that are unique to each acquisition, and a feature that is only prominent in a single acquisition may leak into reconstructions of other acquisitions. Appearance of such artificial features can severely impair diagnostic evaluations; therefore, multi-acquisition reconstructions should be carefully investigated for leakage-of-features.

In this study, we propose a reconstruction method for multi-acquisition MRI, named SIMultaneous use of Individual and joinT regularization terms for joint CS-PI reconstruction (SIMIT). SIMIT leverages both joint and individual regularization terms to maximize sensitivity for shared features among contrasts as well as unique features of each contrast while preventing undesirable leakage-of-features. Specifically, colour TV (CTV) [51] and group $\ell_1$-sparsity [50] are used to exploit common information across contrasts, and individual TV [48] and $\ell_1$-sparsity [47] are used to prevent leakage-of-features. SIMIT is demonstrated for multi-contrast imaging, where the resulting optimization problem is solved efficiently via an adaptation [59] of Alternating Direction Method of Multipliers (ADMM) [27,60,61]. First, SIMIT is compared against alternative reconstructions that only use individual $\ell_1$-sparsity and TV terms (Indiv-only) [59] or only use the joint terms CTV and group $\ell_1$-sparsity (Joint-only) [62], on a numerical phantom dataset. The phantom only included a single-channel receiver coil to isolate potential leakage artefacts. SIMIT is then compared against Indiv-only and Joint-only as well as ESPIRiT reconstructions [63] on multi-channel in vivo datasets. The main contributions of this study are as follows: 1) We introduce the simultaneous use of individual and joint versions of regularization terms in a multi-channel multi-acquisition reconstruction problem. 2) We demonstrate improved image quality and reliability against leakage-of-features in accelerated multi-contrast MRI. Single-channel and multi-channel implementations of this method were presented in part in the 2017 and 2019 Annual Meetings of the International Society for Magnetic Resonance in Medicine [64,65]. In Ref. [64], a single-channel version of the method was presented on a single subject. In Ref. [65], preliminary comparisons were performed between an earlier multi-channel implementation and ESPIRiT. Here, further to Refs. [64,65], we provide a detailed theoretical description of the multi-channel, multi-contrast reconstruction method, thoroughly investigate the benefits of simultaneously using individual and joint regularization terms as opposed to using only joint or only individual terms, demonstrate reconstruction performance across a broad range of acceleration rates and numbers of jointly reconstructed contrasts, and perform qualitative and quantitative comparisons against three state-of-the-art methods for in-vivo data acquired from N=11 participants.

## METHODS

### Theory

We propose to jointly reconstruct multi-contrast datasets by leveraging common information across contrasts via CTV and group sparsity ($\ell_{2,1}$-norm, denoted by $gL1$) regularization while preventing unwanted leakage artefacts via individual TV ($iTV$) and sparsity ($\ell_1$-norm, denoted by $iL1$) regularization. The resulting optimization problem is:

$$\min_x \left[\alpha_{CTV}\; CTV(|\boldsymbol{x}|) + \beta_{gL1}\; \|\boldsymbol{x}\|_{2,1} + \gamma_{iTV}\; \sum_i^k TV\left(\left|\boldsymbol{x}^{(i)}\right|\right) + \theta_{iL1}\; \sum_i^k \left\|\boldsymbol{x}^{(i)}\right\|_1\right] \tag{1}$$

$$\text{subject to } \left\|\boldsymbol{A}^{(i,j)}\boldsymbol{x}^{(i)} - \boldsymbol{y}^{(i,j)}\right\|_2 \le \epsilon_{i,j}, \qquad i \in 1, \ldots, k;\; j \in 1, \ldots, N_c, \tag{2}$$

where $k$ is the number of contrasts, $N_c$ is the number of channels (coils), $\boldsymbol{A}^{(i,j)}$, $\boldsymbol{x}^{(i)}$, $\boldsymbol{y}^{(i,j)}$ and $\epsilon_{i,j}$ denote the encoding matrix, the reconstructed image vector, the received signal acquired through channel $j$ for contrast $i$ and the upper-bound for data-fidelity. Equation (2) denotes the data fidelity constraint for the $i^{\text{th}}$-contrast and $j^{th}$-channel. We prefer including data-fidelity as a constraint as opposed to a Lagrangian form, since $\epsilon_{i,j}$ can simply be set according to the noise level calculated from noise-only data (i.e., data acquired without RF excitation). The CTV, $\ell_{2,1}$, TV, and $\ell_1$ regularization terms can be expressed as:

$$\begin{aligned} CTV(|\boldsymbol{x}|) &= \textstyle\sum_n \sqrt{\sum_{i=1}^k ((\nabla_1 |x^{(i)}[n]|)^2 + (\nabla_2 |x^{(i)}[n]|)^2)}, \\ \|\boldsymbol{x}\|_{2,1} &= \textstyle\sum_n \sqrt{\sum_{i=1}^k |x^{(i)}[n]|^2}, \\ TV(|\boldsymbol{x}^{(i)}|) &= \textstyle\sum_n \sqrt{(\nabla_1 |x^{(i)}[n]|)^2 + (\nabla_2 |x^{(i)}[n]|)^2}, \\ \left\|\boldsymbol{x}^{(i)}\right\|_1 &= \textstyle\sum_n \left|x^{(i)}[n]\right|, \end{aligned} \tag{3}$$

where $\alpha_{CTV}, \beta_{gL1}, \gamma_{iTV}, \theta_{iL1}$ denote the respective regularization parameters. In (3), $\nabla_1, \nabla_2$ denote the image gradients in two orthogonal dimensions. Note that, all functions in (3) can trivially be extended to a higher number of dimensions as would be required for three-dimensional or dynamic acquisitions. The joint regularization terms $\ell_{2,1}$ and CTV combine the contrasts or their spatial derivatives, respectively, before regularizing the images. This allows an image with an inconspicuous tissue boundary that would normally be suppressed via TV regularization to retain this boundary, if the boundary is more prominent in another contrast. However, a prominent but unique feature in one contrast may lead to imperfect noise suppression in other contrasts, leading to misleading artificial tissues. I.e., joint regularization terms enhance reconstructions based on common properties across contrasts but may also lead to leakage of unique features across contrasts. Meanwhile, the individual regularization terms TV and $\ell_1$ suppress interference and noise based on the unique structural properties of each contrast, promising sparse recovery at higher undersampling rates without introducing unwanted feature transfer across contrasts. Therefore, the simultaneous use of the individual terms with the joint terms suppresses leakage-of-features due to the latter. Note that, while the individual and group sparsity terms can be applied in a transform domain that sparsifies the image, they are applied in the image domain in SIMIT, since empirical results in the development stages showed >3 dB better peak signal-to-noise ratio (pSNR) when these terms were applied in the image domain rather than in Wavelet or Discrete Cosine Transform domains. Note that, for typical Wavelet transforms in MRI reconstruction, the $\ell_1$-term in Wavelet domain captures partly similar information to the TV term in image domain. This redundancy is further increased due to simultaneous use of individual and joint versions of regularization terms. In contrast, an $\ell_1$-term in the image domain can improve capture of unique prior information over the TV term. In this study, pSNR was calculated as the ratio of the maximum image intensity across voxels to the root-mean-squared error between the original ($\boldsymbol{x}_0$) and the reconstructed ($\boldsymbol{x}$) images, and expressed in decibels: $\text{pSNR} = 20\log_{10}\frac{\max(|\boldsymbol{x}|)}{\text{RMS}(|\boldsymbol{x}_0-\boldsymbol{x}|)}$.

In the Supporting Information, we give the general ADMM formulation, show how the proposed optimization problem for multi-contrast MRI can be cast in this general formulation, and derive the update rules for implementation. The only parameter not shown in the body of the manuscript, $1/\mu$, is the step-size parameter which determines the rate of convergence; a smaller $\mu$ means larger steps and faster convergence. ADMM is known to converge under mild conditions [66] and the step size should be carefully selected to ensure good convergence behaviour, as the algorithm may diverge for very small $\mu$. An automated way of selecting this parameter is given in [67]. For non-convex problems, if the exact solution of each sub-problem is known, then the algorithm converges to a local minimum.

**Undersampling Masks and Noise**

k-Space data were retrospectively undersampled in one (1D-acceleration) or two (2D-acceleration) phase-encode directions to demonstrate performance for two- and three-dimensional imaging, respectively (Fig. 1). The central one-eighth section of k-space was fully-sampled. For 2D-acceleration, the diameter of the fully-sampled disc was set to one-eighth of the width of the k-space. Sampling masks were generated using probability distribution functions that decayed with a polynomial order of (R-2) or 3 (whichever is larger), where $R$ is the undersampling factor [25]. The undersampling masks were identical across reconstruction methods but were different across contrasts.

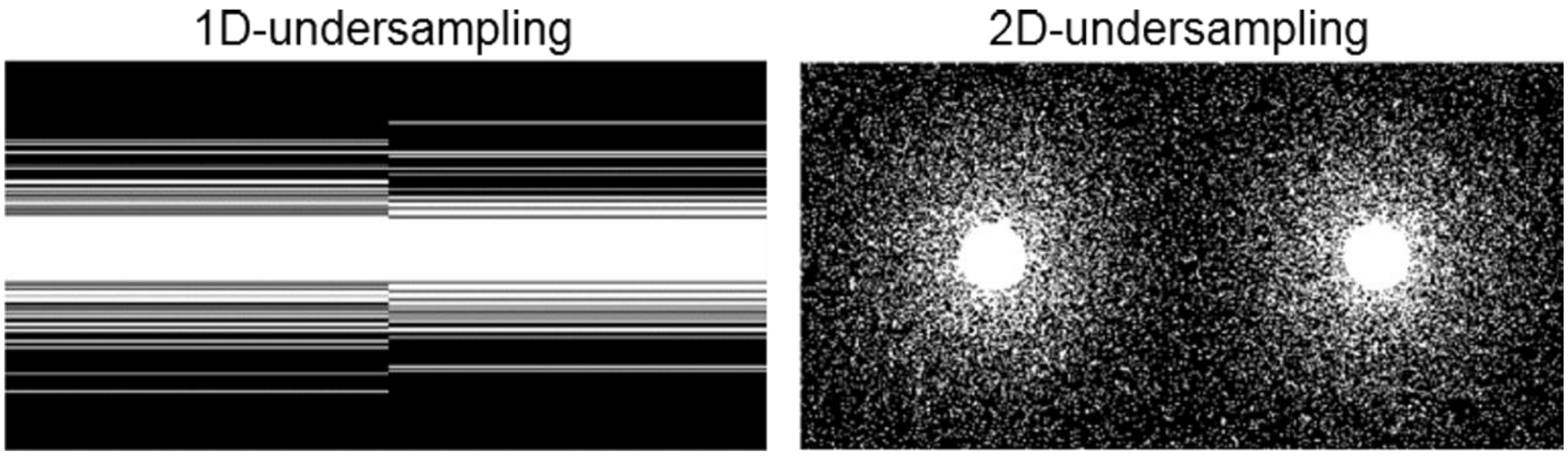

*Fig. 1. Two examples of 1D and 2D-undersampling masks.*

The data-fidelity upper-bounds for SIMIT ($\epsilon_{i,j}$) and its alternative variants (individual terms: Indiv-only; joint terms: Joint-only) were empirically set to half of the square root of the noise power in experimental reconstructions. Note that simulations were designed to investigate different factors that may affect performance. To isolate the effect of such factors on reconstruction performance, noiseless images were used for the simulations, unless specified otherwise.

**Numerical phantom**

The numerical dataset was generated using a realistic individual-subject brain phantom that contained segmentation masks for eleven types of tissues [68]. The original data for the phantom were acquired using a 1.5T scanner with the following parameters: T1-weighted images (3D spoiled gradient echo): TR, 22ms; TE, 9.2ms; flip-angle, 30°; resolution, 1mm isotropic; PD- and T2-weighted images (turbo spin echo): TR, 3300ms; TE (PD/T2), 15/104ms; resolution, 1mm isotropic; number of slices, 62; slice thickness, 2mm; interslice gap, 0mm.

The following contrasts were simulated: PD-weighted (TE/TR: 17/2775 ms), T1-weighted (TE/TR: 14/575 ms), T2-weighted (TE/TR: 102/2775 ms), T1-weighted fluid-attenuated inversion recovery (FLAIR, TE/TI/TR: 17/1050/2775 ms), and T2-weighted short-time inversion recovery (STIR, TE/TI/TR: 17/240/2775 ms). Sinusoidal phase variations in the Anterior-Posterior direction were simulated to introduce image phase, and variations at different spatial frequencies were assumed for separate contrasts (Fig. 2).

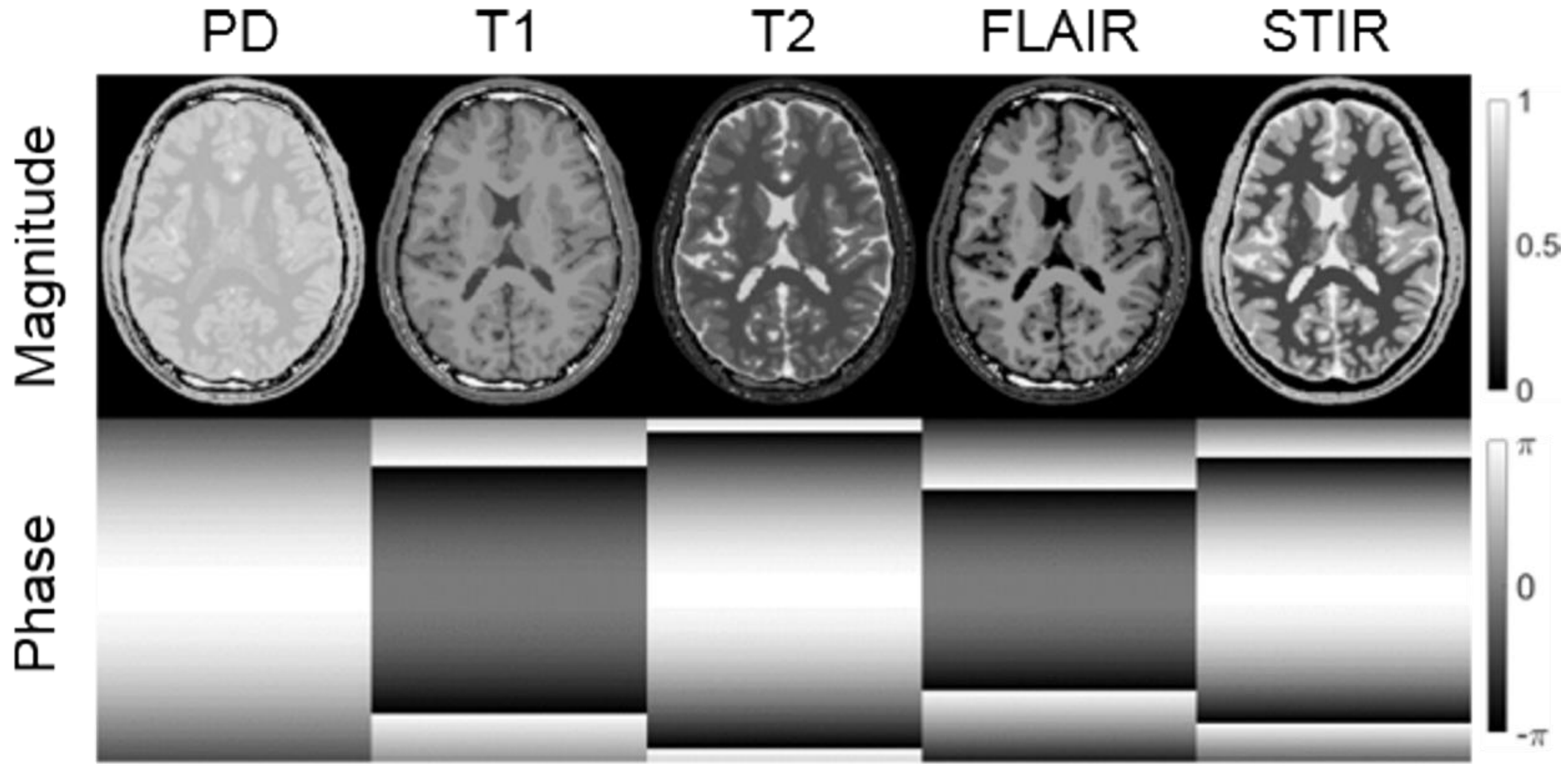

*Fig. 2. Magnitude (top) and phase (bottom) images from the numerical dataset for five different tissue contrasts: proton-density (PD)-weighted, T1-weighted, T2-weighted, fluid-attenuated inversion recovery (FLAIR) and short-TI inversion recovery (STIR) images.*

**In-vivo data**

Comparisons were made using in-vivo raw-data, acquired from N=11 participants using a 3T scanner (Siemens Healthcare, Erlangen, Germany) with a 32-channel receiver-array head coil. Experimental procedures were approved by the local ethics committee and written informed consent was obtained from the participants. A field-of-view (FOV) of 192 mm x 256 mm x 176 mm (phase x readout x slice) and resolution of 1 mm x 1 mm x 2 mm were prescribed for all acquisitions. For T1-weighted acquisitions, an MP-RAGE sequence was used with TE/TI/TR=3.87/1100/2000 ms; flip-angle=20°. For PD- and T2-weighted acquisitions, a turbo spin echo sequence was used with $TE_{PD}$=12 ms, $TE_{T2}$=118 ms, TR=1000 ms, flip-angle=90°; turbo-factor=16, echoes/slice=12. Coil sensitivities were estimated using the approach in [63].

**Parameter Optimization and Image Normalization**

In CS reconstructions, resultant image quality depends on proper selection of regularization parameters (e.g., $\alpha_{CTV}$, $\beta_{gL1}$, $\gamma_{iTV}$, $\theta_{iL1}$, in SIMIT). These parameters are often optimized on held-out training data since fully-sampled data are not available in test subjects. Thus, parameter selection is expected to be suboptimal to varying degrees in practice. To allow for a controlled level of suboptimality, here we intentionally optimized parameters on a five-contrast numerical dataset but tested on datasets with fewer contrasts without dataset-specific optimization.

A mismatch between the image intensities in the training and test datasets may affect reconstruction quality. Here, raw k-space data for each acquisition were normalized such that the images reconstructed using Inverse Fourier Transform (simulations) or ESPIRiT (in-vivo reconstructions) for R=1 spanned the same intensity range, [0,255]. To optimize reconstruction performance, the regularization parameters for SIMIT, Indiv-only and Joint-only were separately optimized to maximize SSIM (structural similarity), which has been suggested to correlate highly with perceptual quality of visual images [69]. Reconstructions were performed with each method for 500 iterations. Each contrast in the five-contrast numerical dataset was 1D-undersampled at R=3. An interval search algorithm was used with grid size: 11x11, depth: 3, parameter range: 0.001 to 2.5. The parameters in [31] were used as initial values and the range was automatically expanded by the algorithm as necessary. The optimized parameters were 0.02 ($\ell_1$-sparsity) and 1.14 (TV) for Indiv-only and 0.085 (Group $\ell_1$-sparsity) and 0.23 (CTV) for Joint-only.

Because the search space for SIMIT with four regularization terms is four-dimensional, the parameters optimized for Joint-only were used for the joint terms and the individual regularization parameters were manually tuned using the joint regularization parameters as the initial values. Note that, the joint regularization terms scale with $\sqrt{k}$ while the sum of the individual terms scales linearly with $k$. Because the parameters were optimized for a five-contrast dataset, parameters were scaled with $\sqrt{5/k}$ and $5/k$. Scaling with $\sqrt{1/k}$ and $1/k$ maintains balance among the regularization terms for an arbitrary number of contrasts $k$, while normalizing the scaling coefficients with $\sqrt{5}$ and 5 keep the coefficients unaltered for $k = 5$, respectively. The optimization yielded $\alpha_{CTV} = 0.19/\sqrt{k}$, $\gamma_{iTV} = 0.11/k$, $\beta_{gL1} = 0.51/\sqrt{k}$, $\theta_{iL1} = 9.13/k$. The convergence speed parameter was empirically selected as $1/\mu = \sqrt{N}/10$, where $N$ is the number of image pixels.

For in vivo experiments, the original pixel intensities in T1-weighted images were approximately an order of magnitude smaller than those in PD- or T2-weighted images. Thus, the normalization of the data scaled the signal in T1-weighted images upwards, preventing a potential mismatch between the image intensities and the regularization parameters. However, because SNR remains constant after normalization, T1-weighted images had higher noise level compared to that of PD- and T2-images during joint reconstructions, leading to noisy reconstructions. To alleviate this issue, parameters for individual regularization terms ($\ell_1$-sparsity and TV) were increased 5-fold for Indiv-only and SIMIT.

An adaptation of ESPIRiT with $\ell_1$-sparsity and TV terms was used to keep regularization terms as consistent as possible across methods under comparison. The parameters were optimized using the same approach as above, but because ESPIRiT uses a fundamentally different algorithm, the initial parameter range was adapted as 0.00005 – 0.125. We observed that optimized parameters yielded over-smoothing in *in-vivo* reconstruction, so the parameters were manually fine-tuned to increase visual acuity and pSNR, yielding 0.00025 ($\ell_1$-sparsity) and 0.000625 (TV). A kernel size of 6x6 was used for ESPIRiT.

**Simulated data**

SIMIT was compared to Indiv-only and Joint-only. All methods used optimized parameters and 250 iterations. To assess reconstruction performance as a function of acceleration rate, methods were compared in terms of pSNR and SSIM for acceleration rates between R=2 and R=5 for 1D-acceleration and between R=4 and R=15 for 2D-acceleration.

To investigate performance as a function of the number of acquisitions that are jointly reconstructed, SIMIT was performed on 2D-accelerated data with undersampling factors between R=4 and R=15. The number of contrasts was varied from 1 to 5, and for each number all possible subsets of the five-contrast dataset were considered. pSNR and SSIM were averaged across 10 initializations of the undersampling mask for each case. pSNR and SSIM were also averaged across all subsets that contained a given contrast, e.g., the SSIM of the PD image was averaged across PD-T1, PD-T2, PD-FLAIR and PD-STIR for two-contrast joint reconstruction.

Methods were also compared in terms of reconstruction time and stability of performance across undersampling masks and noise instances via a Monte-Carlo simulation with 250 runs. The five-contrast dataset was used with 1D-undersampling and R=3. Each run was performed with independent instances of noise and undersampling masks. Runtimes (excluding data-preparation) at each iteration were measured with the cputime command in Matlab (which excludes any parallel computing capabilities) and averaged across runs. SSIM and pSNR averaged across contrasts and runs were plotted as a function of cumulative runtime for each method. Bivariate Gaussian noise was added with a standard deviation equal to 10% of the mean intensity of k-space data across all contrasts. A second set of Monte-Carlo tests were conducted (25 runs) to investigate the effect of using the same undersampling masks for each contrast versus using different masks across contrasts.

An important concern regarding the use of joint regularization terms is leakage-of-features that are unique to a subset of the contrasts to other contrasts. To assess reliability against leakage-of-features, an elliptical dark region was introduced artificially in the PD-weighted image and an elliptical bright region was introduced in the T1-weighted image. These regions did not overlap spatially, and their intensities were accordingly set to the minimum and the maximum of their respective images. To increase the potential effect of unique features on the joint regularization terms, the dataset was reduced to three contrasts (PD-, T1-, T2-weighted). All acquisitions were 2D-accelerated with R=4.

To test stability against variations in the regularization parameters, all four parameters in SIMIT ($\alpha_{CTV}$, $\beta_{gL1}$, $\gamma_{iTV}$, $\theta_{iL1}$) were individually scaled up/down until average SSIM across images (PD-, T1-, T2-weighted; 2D-acceleration, R=3) decreased from 98% (optimized parameters) to below 95%. Since $\mu$ primarily controls convergence rate, it was not altered.

**In-vivo data**

SIMIT was compared to Indiv-only and Joint-only as well as the state-of-the-art ESPIRiT method for retrospectively 2D-undersampled *in-vivo* data from N=11 participants. Methods were compared in terms of pSNR and SSIM (R=8, R=12, R=16) as well as via neuroradiologist reader studies (R=8). Images reconstructed with ESPIRiT without any undersampling (R=1) were used as reference in both evaluations. A neuroradiologist reader with 18 years of experience was blinded to method names, and methods were presented in randomized order. The reader evaluated the images for anatomical detail (1: low, 2: fair, 3: good/ acceptable for clinical use, 4: very good, 5: excellent) as well as Gibbs artefacts and noise level (1: intolerable, 2: too much, 3: acceptable/ not degrading the image, 4: very little, 5: none). All reference images were assigned a score of 5 in all categories by the reader to set a benchmark. Wilcoxon signed-rank test was performed on the reader scores as well as pSNR and SSIM measurements.

**Further Comparisons (Supporting Information)**

The proposed method was also compared to a collection of 6 state-of-the-art individual and joint reconstruction methods from the literature; Sparse MRI [25], TVCMRI [26], RecPF [27], GSMRI [29], FCSA [28] and FCSA-MT [31] using the numerical phantom and in-vivo data from a participant. Because these reference methods are for a single-coil receiver coil, these comparisons [65] are given in the Supporting Information for conciseness.

## RESULTS

**Simulation Results**

SIMIT consistently outperformed both Indiv-only and Joint-only in terms of SSIM for 1D- and 2D-acceleration, demonstrating the benefit of simultaneously using individual and joint regularization terms (Fig. 3). On the average, SIMIT improved pSNR compared to Indiv-only and Joint-only by 1.7 dB and 4 dB for 1D-acceleration, and by 1.7 dB and 4.4 dB for 2D-acceleration, respectively. While the difference in SSIM is rather small for two-fold 1D-acceleration, the difference between the methods becomes evident as the acceleration factor increases, with SIMIT yielding 1.6% and 3.6% better SSIM than Indiv-only and Joint-only, respectively. For the range of 2D-acceleration factors shown, SIMIT yields 2.1% and 5% better SSIM compared to Indiv-only and Joint-only, respectively.

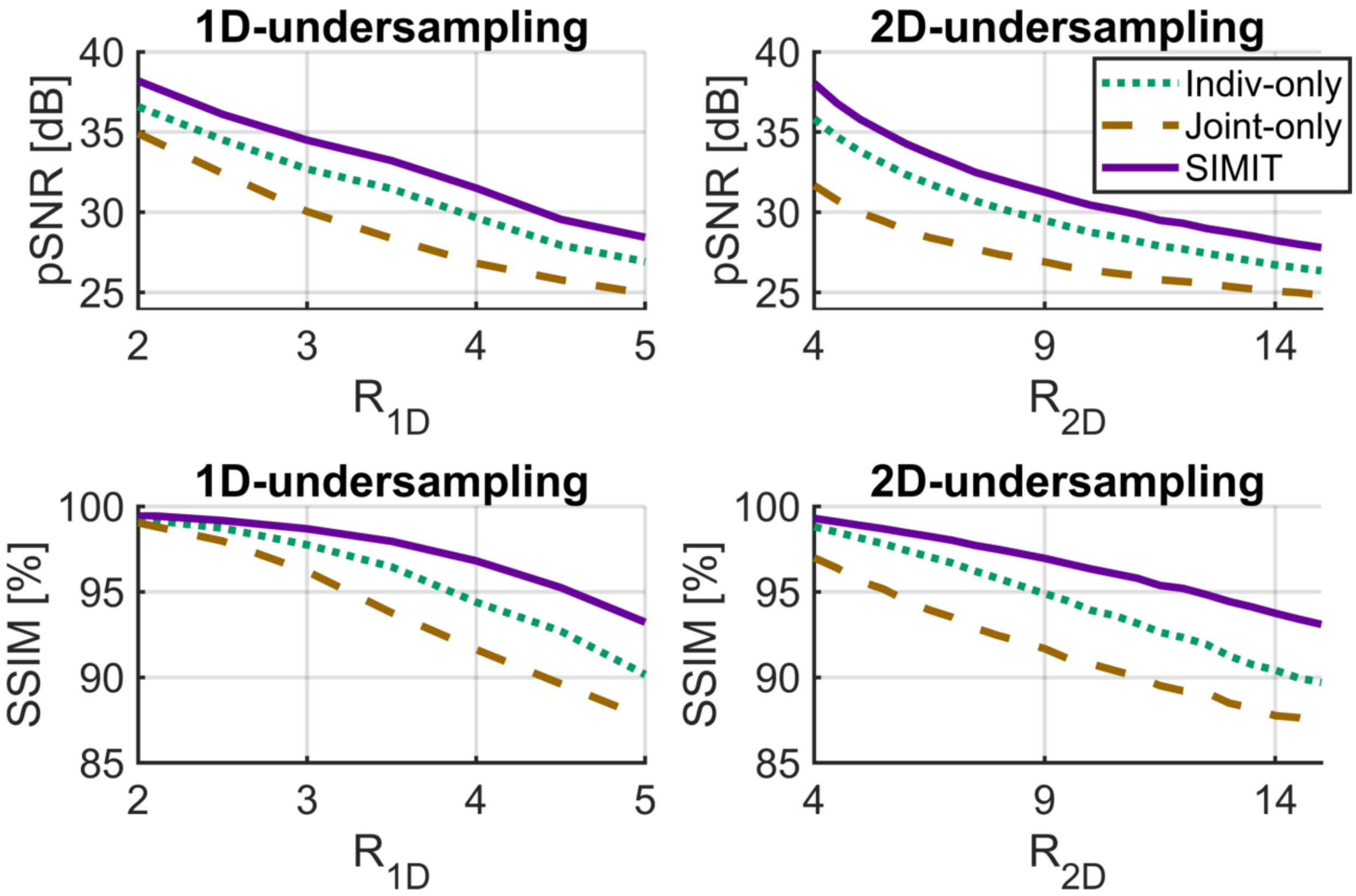

*Fig. 3. Simultaneous use of individual and joint regularization terms (SIMIT) improves image quality (in terms of pSNR and SSIM) over using only individual (Indiv-only) or only joint (Joint-only) regularization terms at all examined acceleration factors for both one-dimensional ($R_{1D}$=2 to 5), and two-dimensional ($R_{2D}$=4 to 15) acceleration. SSIM and pSNR were averaged across contrasts.*

Different contrasts have different pSNR and SSIM values for the same acceleration factors as pSNR and SSIM depend on image content (Fig. SI-8 in Supporting Information). Nevertheless, regardless of their individual pSNR and SSIM levels, Fig. 4 demonstrates that all contrasts benefit from joint reconstruction, and consistently across the examined range of acceleration factors between R=4 and R=15. The performance improvement with increasing number of contrasts is more noticeable for higher acceleration factors. Averaged across contrasts, SSIM and pSNR were 0.8% and 2 dB higher for 5-contrast reconstruction than individual reconstruction for R=4. For lower acceleration rates, SSIM curves are saturated near 100% and pSNR performance is enhanced for greater number of contrasts. For higher accelerations, PSNR curves are relatively flat and SSIM performance is enhanced towards higher number of contrasts. Comparing a 5-contrast reconstruction to individual reconstruction, SSIM improvement monotonically increases to 3.3% SSIM at R=15, while the pSNR improvement monotonically decreases to 1 dB at R=15.

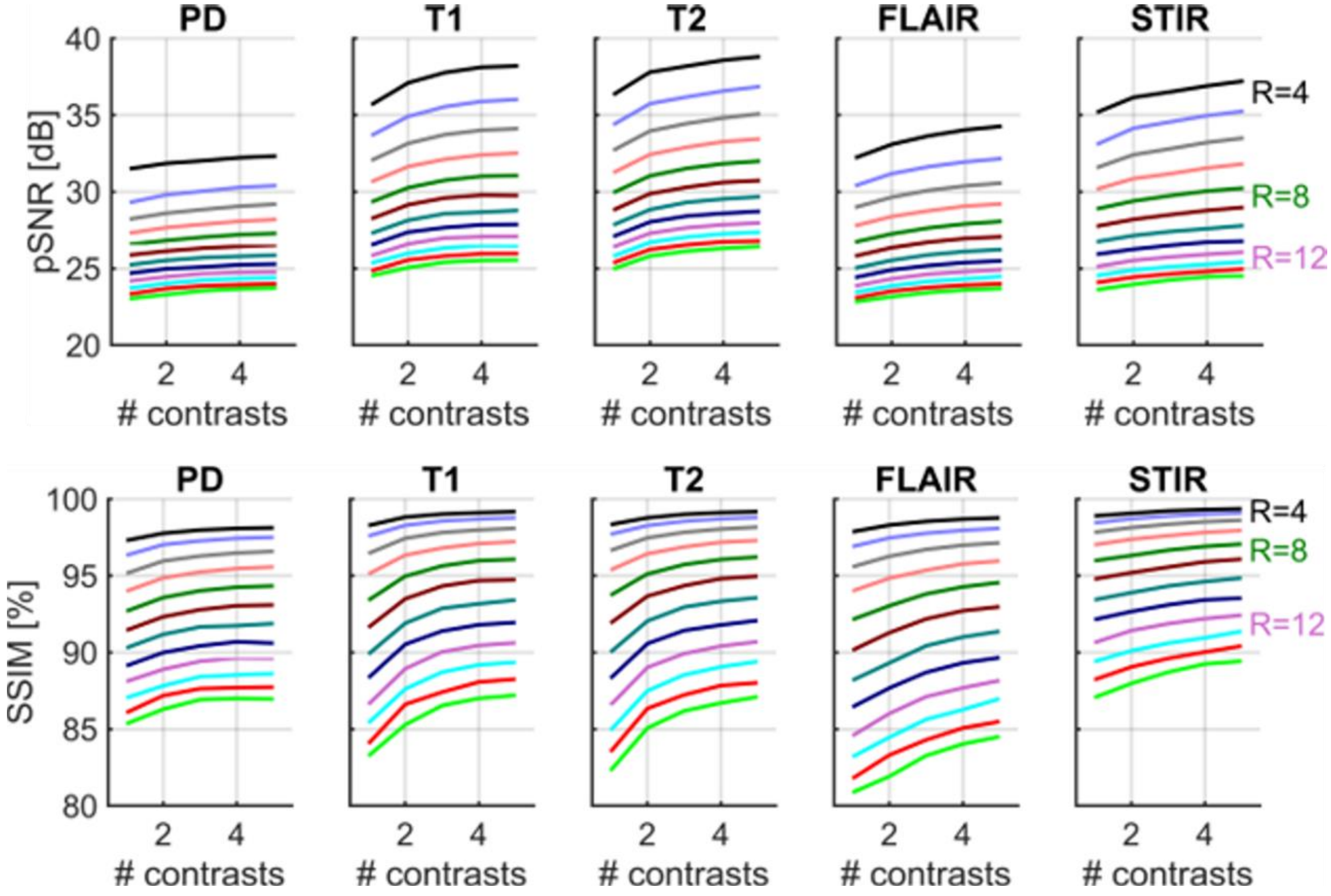

*Fig. 4. The variations of pSNR and SSIM with respect to the number of jointly reconstructed contrasts and the two-dimensional acceleration factor R is shown for SIMIT. As the number of contrasts that are jointly reconstructed increases, image quality increases for all contrasts and acceleration factors. For a given contrast and a given number of jointly reconstructed contrasts, pSNR and SSIM were averaged across all possible subsets of contrasts, e.g. for PD and 2-contrast reconstruction, SSIM of the PD-image was averaged across PD-T1, PD-T2, PD-STIR and PD-FLAIR reconstructions.*

When identical undersampling masks were used for all contrasts instead of different masks across contrasts, the pSNR values dropped by up to 0.1 dB across contrasts (data not shown). Therefore, the effect of using different versus identical undersampling masks on the performance of the proposed method was not further considered.

Despite increases in computation time per iteration with more regularization terms in SIMIT, the simultaneous use of individual and joint regularization terms in SIMIT enables improved reconstruction performance. Fig. 5 shows the variations of pSNR and SSIM with respect to reconstruction time, and variations in undersampling masks and noise instances. SIMIT quickly surpasses Indiv-only and Joint-only to yield higher pSNR and SSIM before Indiv-only and Joint-only can converge to their final images. Note that these computation times exclude any parallel computation. Fig. 5b shows the change in SSIM and pSNR due to variations in noise and masks. SIMIT is less sensitive to these variations, yielding similar or lower standard variation across the Monte-Carlo runs.

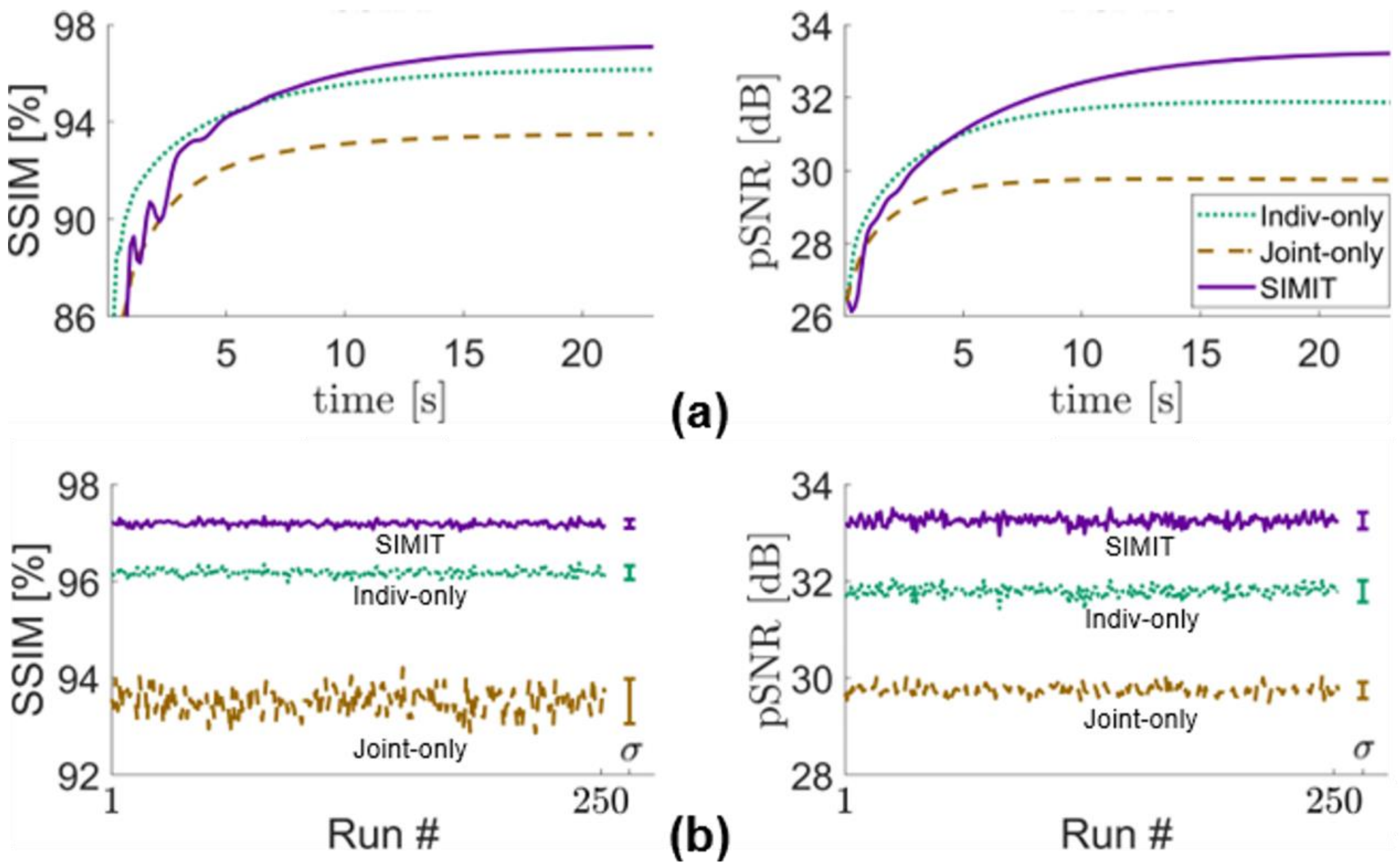

*Fig. 5. The variations of pSNR and SSIM with respect to **(a)** the reconstruction time and **(b)** variations in noise instances and undersampling masks. SSIM and pSNR were averaged across contrasts. In panel (a), SSIM and pSNR were averaged over 250 Monte-Carlo runs. Although SIMIT is mildly slower in improving image quality metrics in the first iterations, it quickly surpasses Indiv-only and Joint-only, yielding higher quality images before the metrics can reach steady-state for Indiv-only and Joint-only. In (b), SIMIT shows similar or better stability against variations in noise instances and undersampling masks. Standard deviation values are Indiv-only: 0.14% and 0.22dB, Joint-only: 0.46% and 0.17dB, SIMIT: 0.09% and 0.17 dB for SSIM and pSNR, respectively.*

Fig. 6 visually compares SIMIT, Indiv-only and Joint-only in terms of leakage-of-features and reconstruction artefacts. Although Indiv-only does not have any leakage since images are reconstructed separately, the images suffer from residual noise-like artefacts due to undersampling. Joint-only reduces these undersampling artefacts but suffers from two potential drawbacks of joint reconstruction; leakage-of-features are apparent in all images (red arrows) and blurring of unique features are seen in the dark crescent in the original T1-weighted image (green arrow). SIMIT does not have any structured leakage-of-features, and the intensity of the image reconstruction artefacts due to undersampling are below those observed for Indiv-only and Joint-only. Furthermore, the dark crescent is as clearly represented as in Indiv-only. These results indicate that the simultaneous use of individual and joint terms prevent both potential pitfalls of joint reconstruction. SIMIT also alleviates the staircase artefacts seen for Joint-only and suppresses the noise-like artefacts seen for Indiv-only.

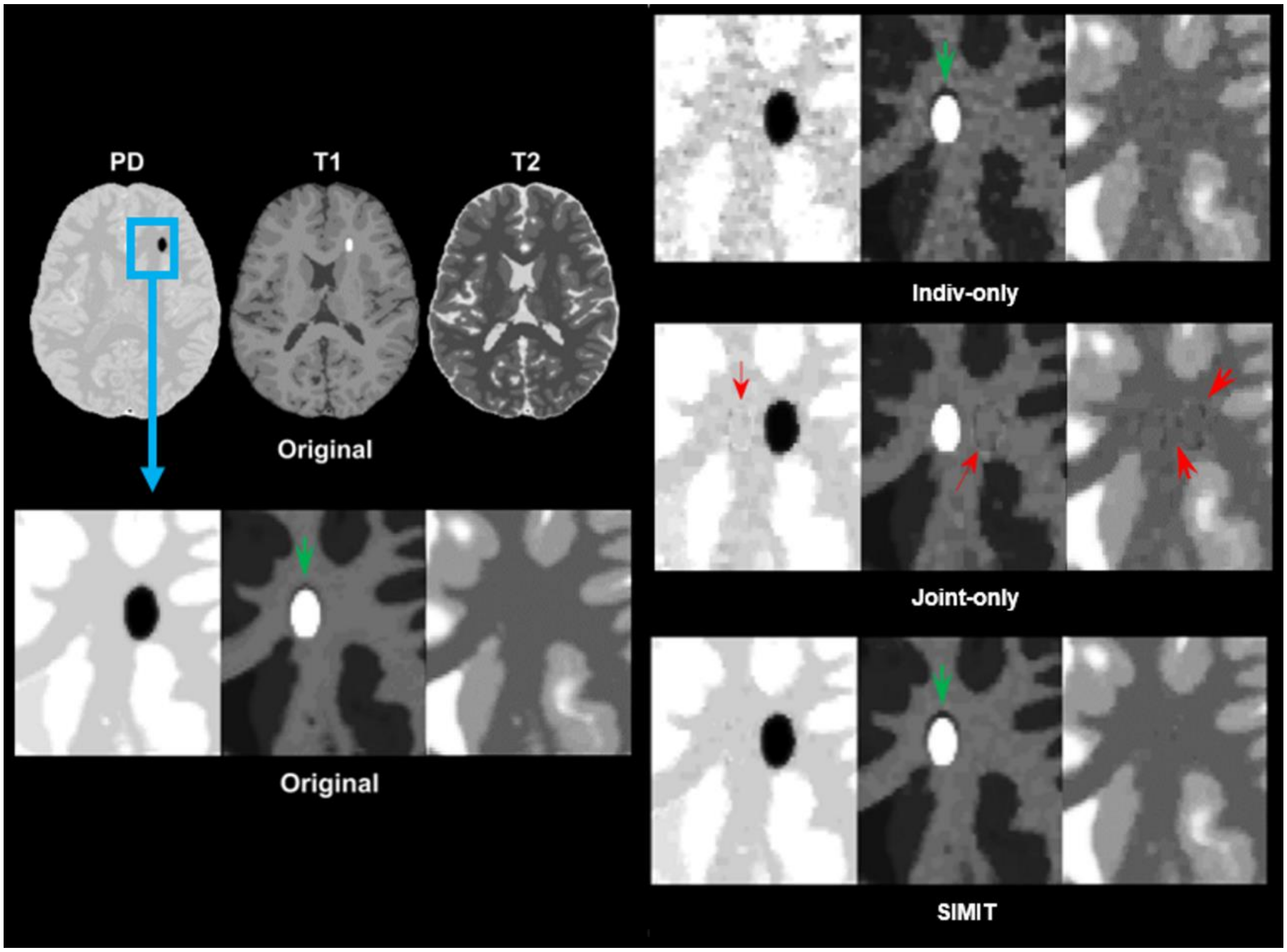

*Fig. 6. SIMIT, Indiv-only and Joint-only were compared in terms of leakage-of-features across contrasts. The numerical phantom was stripped of the skull and the skin for visualization purposes (but both tissues were included in the simulations). Red arrows show the leakage-of-features in Joint-only reconstruction, which are suppressed with SIMIT. Green arrow shows that the black crescent seen in the original image is blurred in Joint-only, but it is clearly delineated with SIMIT.*

With optimized regularization parameters, SIMIT yields an average SSIM of 98.1% for the 3-fold 2D-accelerated three-contrast dataset. Individually scaling down $\alpha_{CTV}$, $\gamma_{iTV}$, $\beta_{gL1}$, $\theta_{iL1}$ up to an order of magnitude did not reduce SSIM below 95%. Scaling the parameters up makes the regularization functions penalize the reconstruction more heavily and may lead to suboptimal reconstruction performance. SSIM still remained above 95% when $\alpha_{CTV}$, $\gamma_{iTV}$ and $\theta_{iL1}$ were individually scaled up by an order-of-magnitude. However, although SSIM remained above 95% when $\beta_{gL1}$ was scaled up 5-fold, it reduced below 95% when $\beta_{gL1}$ was further doubled. All parameters except $\beta_{gL1}$ have an order-of-magnitude headroom downwards and upwards, before noticeably affecting the image quality.

**In-vivo results**

SIMIT, Indiv-only and Joint-only as well as ESPIRiT were compared on in-vivo multi-channel acquisitions from N=11 participants. Magnified regions-of-interest (ROI) from representative reconstructions are shown in Fig. 7 for PD-, T1- and T2-weighted images at R=8. Visual comparisons show that with respect to Indiv-only and Joint-only, which suffer from residual reconstruction error and noise, SIMIT yields better noise suppression and leads to a clearer depiction of tissues, particularly manifested inside the Lentiform Nucleus and the Putamen in the higher-SNR PD- and T2-weighted images (Figs. 7a and 7c). Furthermore, the grey-matter and white-matter boundaries are hard to distinguish in the lower-SNR T1-weighted image for Indiv-only and Joint-only, whereas SIMIT yields a clear depiction of these boundaries (Fig. 7b). ESPIRiT also yields better noise suppression than Indiv-only and Joint-only, albeit at the cost of blurring and Gibbs-artefacts. Compared to ESPIRiT, SIMIT yields sharper images with better suppression of Gibbs-artefacts. This leads to a more accurate representation of the Globus Pallidus in the PD-weighted image, better delineation of the grey- and white-matter boundaries in the T1-weighted image, and the Putamen in the T2-weighted image.

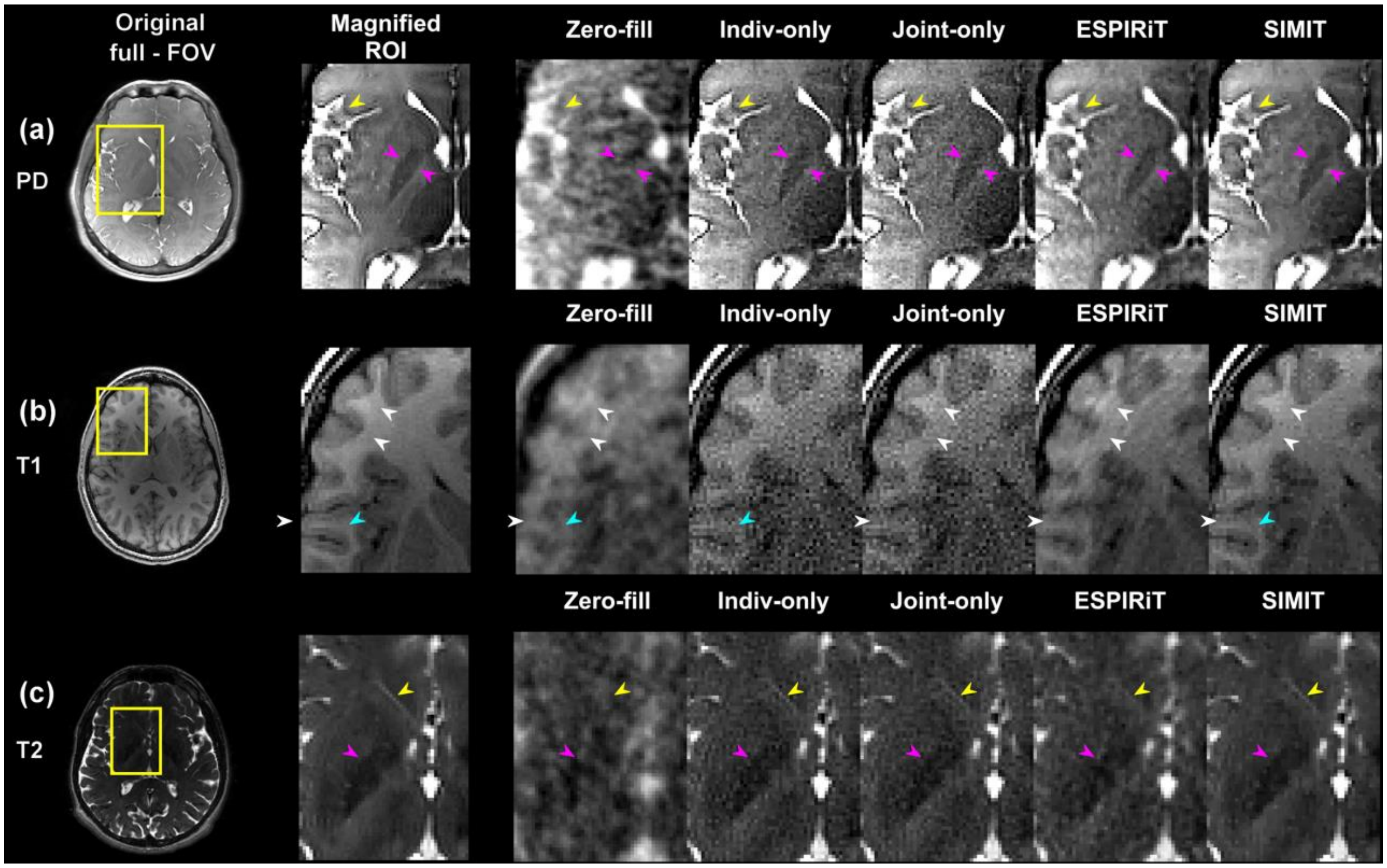

*Fig. 7. Representative reconstructions of* ***(a)*** *PD-weighted,* ***(b)*** *T1-weighted,* ***(c)*** *T2-weighted images from three different participants for all methods at R=8. Magnified views show the regions bounded by the yellow rectangles.* ***(a)*** *Indiv-only and Joint-only suffer from noise, while ESPIRiT shows blurring at the boundary of the frontal opercular cortex (yellow arrows) and a narrower representation of the Globus Pallidus (pink arrows). SIMIT yields better noise suppression while demonstrating a clearer delineation of tissue at the frontal opercular cortex (yellow*

*arrow) and inside the Lentiform Nucleus (pink arrows). **(b)** Due to the relatively lower SNR of the T1-weighted image, grey-matter boundaries in the sulci cannot be identified in Indiv-only and Joint-only reconstructions (cyan arrow). For ESPIRiT, while noise suppression is much better, tissue delineation is compromised in the gyri due to the Gibbs-like artefacts (white arrows). SIMIT yields much better suppression than Indiv-only and Joint-only while yielding a clear depiction of grey-matter white-matter boundaries without Gibbs-artefacts. **(c)** SIMIT yields better delineation of the Putamen (pink arrow) as well as the partial volume of the Lateral Ventricle (yellow arrow) compared to the other methods.*

Indiv-only and Joint-only show elevated levels of noise-like error while ESPIRiT and SIMIT yield improved suppression of noise-like artefacts as demonstrated by the error images between the reconstructions and the ideal reference of a representative participant in Fig. 8 (R=8). While the intensities of the error images are similar for ESPIRiT and SIMIT in the lower-SNR T1-weighted image, SIMIT outperforms ESPIRiT in artefact suppression for the higher-SNR PD- and T2-weighted images.

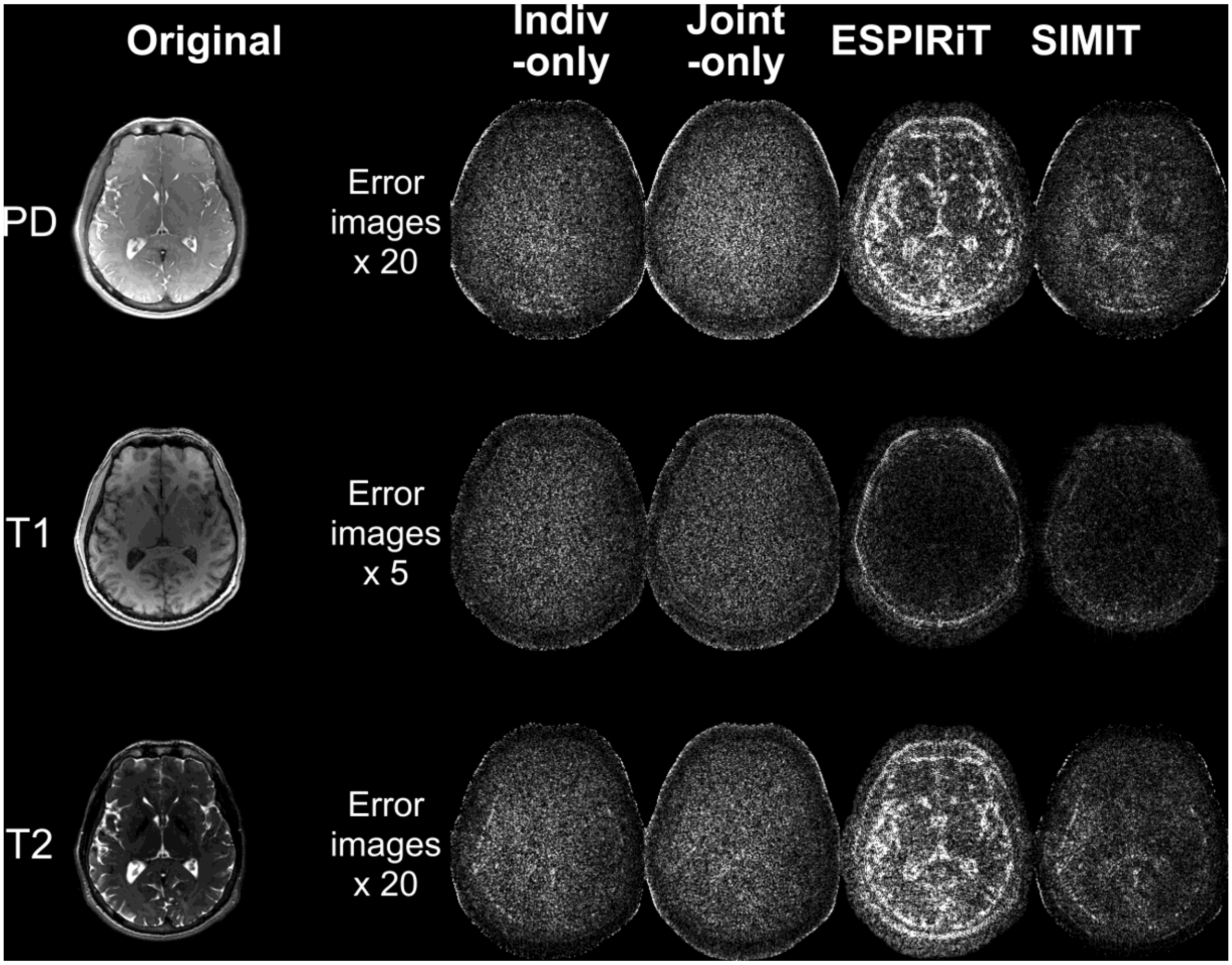

*Fig. 8. Error images were calculated between the fully-sampled reference image and the reconstructed images for all methods at R=8. Indiv-only and Joint-only suffer from noise-like reconstruction artefacts for all contrasts. For the lower-SNR T1-weighted image, the error images for ESPIRiT and SIMIT have similar intensity, although the error for ESPIRiT is considerably more intense than SIMIT for PD- and T2-weighted images.*

The error images were summed across all participants and contrasts to compare the methods for different acceleration factors (Fig. 9). The reconstruction artefacts for SIMIT are visually less intense for all acceleration factors. Sparse reconstructions contain both noise-like and structured artefacts due to undersampling. In visualization of reconstructed images, the reduced intensity of noise-like artefacts in SIMIT might give the impression that structured artefacts are more prominent compared to Indiv-only and Joint-only, even though the latter methods also have similar levels of structured artefacts. To investigate this issue, we calculated the error images between each reconstructed image and the fully-sampled reference image. The error images for separate reconstruction methods were then subtracted from each other to demonstrate potential differences in artefacts (Fig. 10). The difference images have only noise-like behaviour and lack any structural information. This confirms that Indiv-only and Joint-only have similar levels of structured artefacts as SIMIT, but these are overshadowed by the higher levels of noise-like error in Indiv-only and Joint-only.

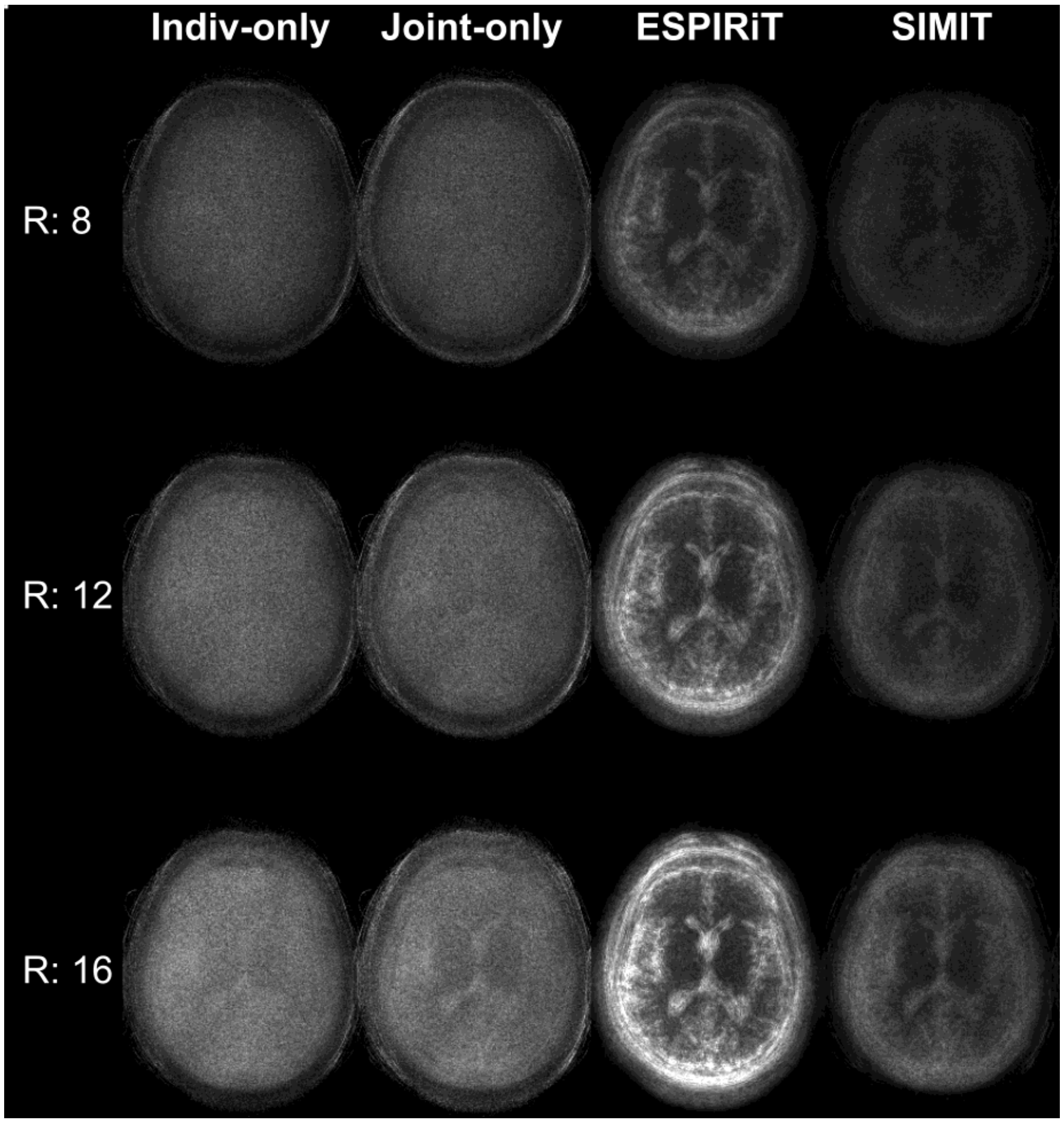

*Fig. 9. Maps of reconstruction error were calculated for each contrast in each individual subject. Error maps were averaged across all contrasts and participants. Error maps were intensified 10-fold and shown in the same colour-scale as the originals in Fig. 8. Error maps are shown for all methods at R=8, R=12 and R=16. On average, SIMIT yields visually reduced reconstruction artefacts compared to reference methods.*

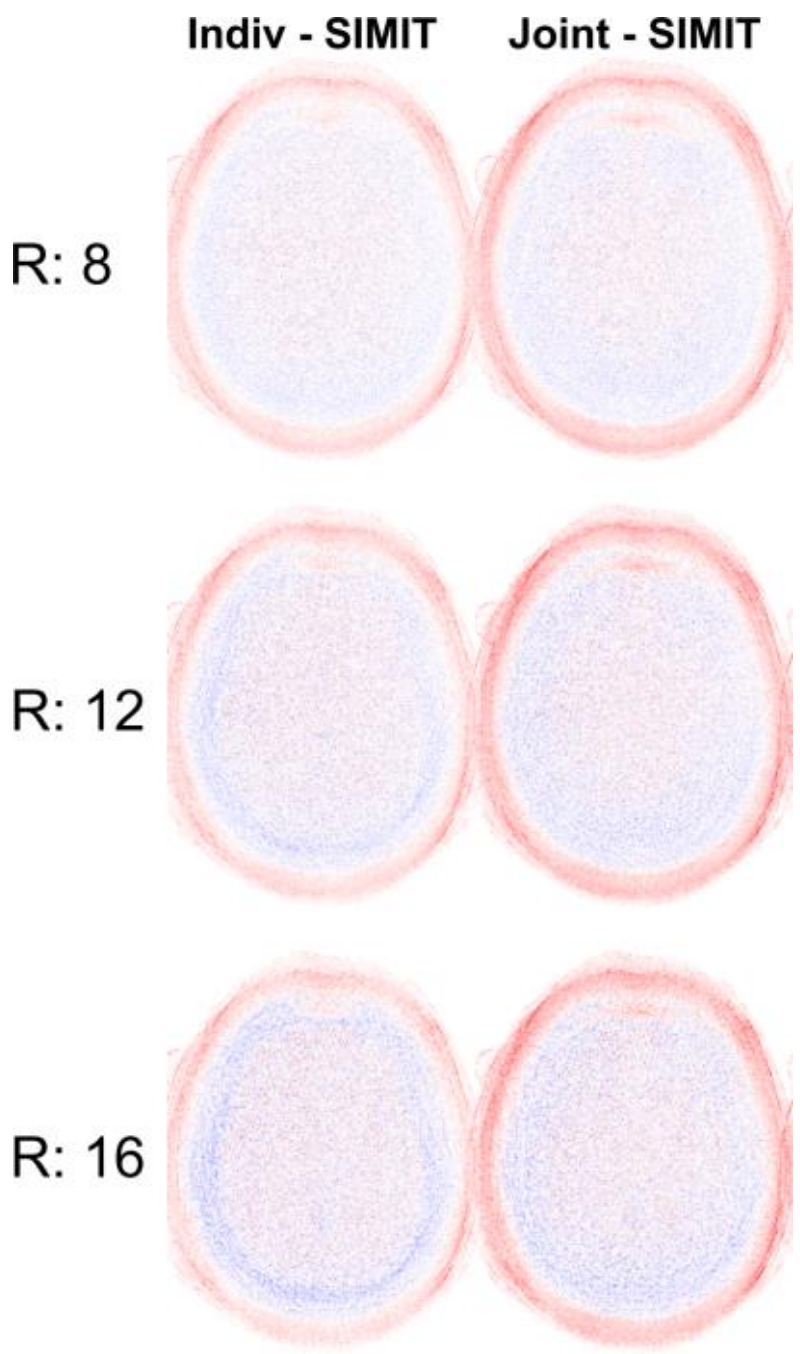

*Fig. 10. The difference images for SIMIT were subtracted from the difference images for Indiv-only and Joint-only, and then summed over N=11 participants. The resulting maps show noise-like behaviour but lack structural information. This demonstrates that structured artefacts are at similar levels for SIMIT, Indiv-only and Joint-only.*

Methods were also compared quantitatively. Statistical analysis showed that SIMIT yields significantly better pSNR and SSIM values for all contrasts and acceleration factors (R=8, R=12 and R=16) compared to all methods ($p<0.05$, Fig. 11). Averaged across participants, contrasts and acceleration factors, SIMIT yielded higher pSNR values than Indiv-only, Joint-only and ESPIRiT by 4.5 dB, 5.0 dB and 4.3 dB at R=8; by 4.1 dB, 4.4 dB and 6.0 dB at R=12; and by 3.6 dB, 3.7 dB and 6.5 dB at R=16, respectively. Compared to all reference methods across all acceleration factors, SIMIT yielded at least 3.6 dB improvement in pSNR.

ESPIRiT yielded relatively more consistent reconstruction performance across contrasts, with less than 3 dB variation in (participant-averaged) pSNR across contrasts for all acceleration factors. For Indiv-only and Joint-only, the variation across contrasts was as high as 10 dB. SIMIT yielded more consistent results than Indiv-only and Joint-only with up to 6 dB pSNR variation across contrasts. Even though the variation was larger than that of ESPIRiT, comparing the maximum pSNR values of ESPIRiT (blue triangular markers) and the minimum pSNR values of SIMIT (red triangular markers) shows that the pSNR values of SIMIT were higher than those of ESPIRiT for all contrasts, acceleration factors and participants.

Joint reconstruction via SIMIT allows increasing the acceleration factor without compromising image quality. For the PD-weighted image, SIMIT allows increasing R=8 to R=12 compared to Indiv-only and ESPIRiT, and R=16 compared to Joint-only while improving pSNR and SSIM. For the T1-weighted image, R=8 can be increased to R=10 (not shown) compared to ESPIRiT and R=16 compared to Indiv-only and Joint-only with better pSNR and SSIM. For the T2-weighted image, SIMIT yields better pSNR and SSIM at R=10 than Indiv-only and Joint-only at R=8, and at R=16 than ESPIRiT at R=8.

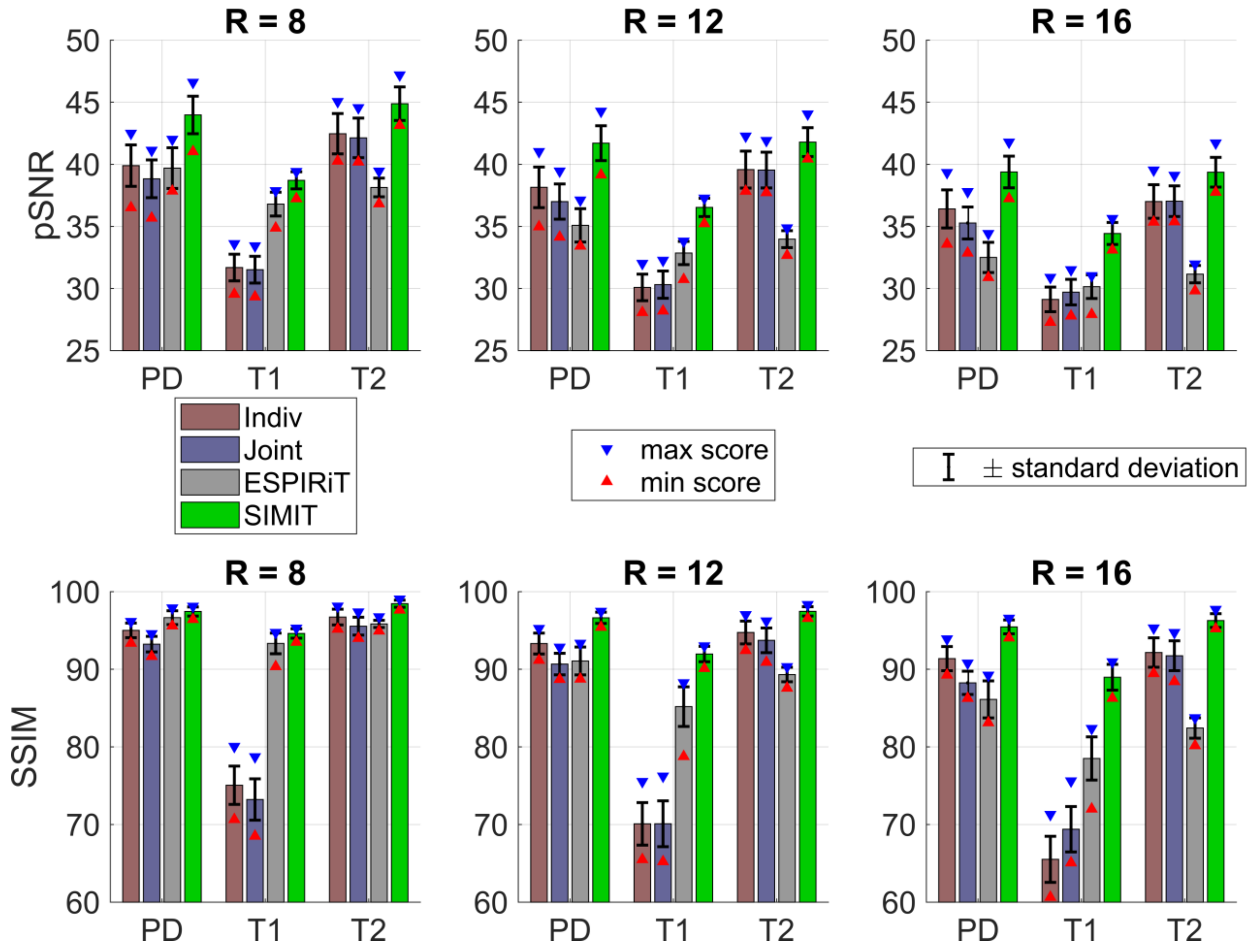

*Fig. 11. Methods are compared in terms of pSNR and SSIM for all participants and contrasts at R=8, R=12 and R=16. SIMIT yields significantly higher pSNR and SSIM (p<0.05) than all methods, consistently across acceleration factors and contrasts. Blue and red arrows show the maximum and minimum values, respectively, and the error bars show the standard deviation of the measured metric (pSNR or SSIM).*

To confirm that the visual and quantitative improvements in image quality enabled by SIMIT translate to diagnostic assessment, neuroradiologist reader studies were conducted for R=8 (Fig. 12). For all contrasts and comparisons in terms of anatomy, noise and Gibbs-artefacts, SIMIT yields higher scores than the other methods. SIMIT yields significantly better ($p<0.05$) anatomy scores than the other methods except for T1-weighted against ESPIRiT, where the two methods perform similarly. In terms of noise, SIMIT performs significantly better than Indiv-only and ESPIRiT for two of the contrasts while performing similarly for a third, and it performs significantly better than Joint-only for all contrasts. In terms of Gibbs artefacts, SIMIT performs significantly better than all other methods for PD-weighted images. SIMIT performs significantly better than ESPIRiT for T1-weighted images, while performing similarly to Indiv-only and Joint-only. Meanwhile, all methods perform similarly for T2-weighted images. For each contrast, SIMIT yields significantly better scores ($p<0.05$) in at least one of the comparisons (anatomy/noise/Gibbs) against each alternative method.

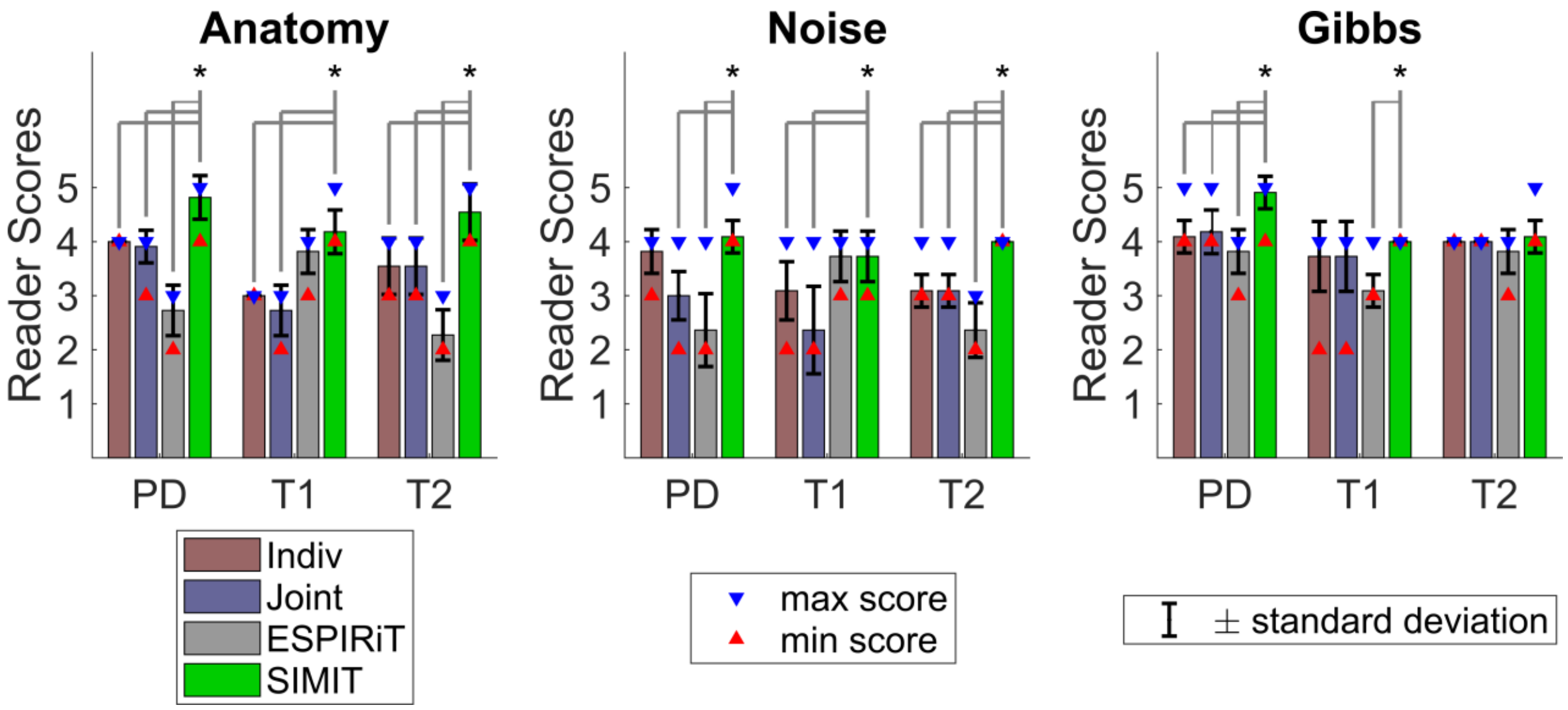

*Fig. 12. Reconstruction methods were compared in terms of neuroradiologist reader scores. The reader was blinded to method names and methods were presented in randomized order. SIMIT yields significantly higher scores in 19 out of 27 comparisons and yields similar performance in the remaining cases. The methods SIMIT yields significantly higher scores against are indicated by the asterisks and the vertical bars below the asterisks (e.g. against Joint-only and ESPIRiT for the T1-weighted image in terms of noise-level). Blue and red arrows show the maximum and minimum scores, respectively, and the error bars show the standard deviation of the scores.*

SIMIT performs significantly better ($p<0.05$) than Indiv-only, Joint-only and ESPIRiT in six, seven and six out of nine comparisons (three contrasts, anatomy/noise/Gibbs), respectively. Averaged across participants, contrasts and categories, the neuroradiologist scores for SIMIT were higher by 0.7, 0.9 and 1.2 compared to Indiv-only, Joint-only and ESPIRiT, respectively.

**Further Comparisons (Supporting Information)**

SIMIT was also compared to a large collection of state-of-the-art methods from the literature [25-29,31,59] in terms of pSNR, SSIM and computation speed (Supporting Information). While the reference method RecPF had a faster initial improvement in pSNR and SSIM, SIMIT quickly surpassed RecPF and converged to higher quality images (Fig. SI-2). SIMIT was also more robust to variations in undersampling masks and noise (Fig. SI-2), yielded better suppression of reconstruction artefacts (Figs. SI-3 and SI-4), and provided better SSIM and pSNR for various 1D- and 2D-acceleration rates (Fig. SI-5). Finally, SIMIT yielded a clearer and sharper depiction of tissues for the in-vivo data (Figs. SI-6 and SI-7).

## DISCUSSION

The proposed multi-channel multi-acquisition reconstruction method, SIMIT, incorporates both joint and individual regularization terms across multi-contrast images. The complex optimization problem that arises is solved via the ADMM algorithm. SIMIT enhances sparse recovery for multi-contrast datasets, for both single- and multi-channel receiver coils. It also enables prescription of higher acceleration factors through joint reconstruction of multi-contrast acquisitions. In multi-contrast reconstructions, SIMIT outperforms a variant method that only includes individual regularization terms (Indiv-only), a variant that only includes joint regularization terms (Joint-only), as well as a state-of-the-art parallel imaging method (ESPIRiT). Compared to Indiv-only and Joint-only, SIMIT lowers reconstructions errors due to residual noise and aliasing. While Joint-only suffers from visible feature leakage across contrast, SIMIT yields enhanced reliability against these artefacts. SIMIT also improves recovery of high spatial frequency details compared to ESPIRiT. The enhanced image quality of SIMIT is also apparent in both quantitative metrics and neuroradiologist reader scores.

Even though the proposed method uses four regularization terms, the additional time required by using more regularization terms is relatively small compared to the time required to complete a whole iteration. Furthermore, using all terms simultaneously improves image quality with respect to individual-only and joint-only reconstruction due to information sharing and prevention of leakage-of-features, respectively. Therefore, image quality rapidly improves in fewer iterations, and the method converges to higher-quality images.

In this study, non-identical undersampling masks were used for each contrast. Even though random undersampling patterns in CS lead to incoherent undersampling artefacts, using identical masks for each contrast may create a coherence in the undersampling artefacts across contrasts. The method was also tested with identical undersampling masks across contrasts, however, the difference with respect to using non-identical masks was modest. This could partly be attributed to the dissimilarity of the jointly reconstructed contrasts. Even though the undersampled frequencies are the same, the energy content at these frequencies is different for each contrast, leading to dissimilar undersampling artefacts. Using identical undersampling masks could potentially lead to further reductions in imaging performance for similar contrasts such as multi-echo acquisitions at different echo-times. Detailed comparison of using identical undersampling masks to undersampling masks explicitly designed to complement each other can be found in Ref. [43,70].

Previous joint reconstruction approaches in MRI include using nuclear and Frobenius norms [35,36] for dynamic MRI; K-SVD [71] for parametric mapping [37]; minimizing the sum of individual regularization functions [38], spatially weighting the regularization terms of an image using a prior image for multi-contrast MRI [49], and replacing one or both of the $\ell_1$-sparsity and TV terms with group sparsity and CTV for diffusion tensor imaging [39], parametric mapping [40], multi-echo T2-weighted imaging [29,33] and multi-contrast imaging [30,31]. In this study, our choice in regularization terms was motivated by two reasons. First, we preferred the more commonly used $\ell_1$-sparsity and Total Variation to other alternatives, since specific terms used in dynamic MRI and parametric mapping may not be directly applicable to multi-contrast datasets that comprise a small number of static acquisitions under distinct contrasts, and that may not lead to an overcomplete dictionary suitable for K-SVD. Second, we simultaneously used individual and joint versions of the regularization terms to create a balance between utilizing common features across images and preserving individual features of each contrast. Group $\ell_1$-sparsity was introduced for improving signal recovery in low-SNR voxels, in cases where the signal is present broadly across contrasts. Note that group-sparsity can also lead to unwanted suppression of a signal that is present only in a small subset of contrasts. For such cases, individual sparsity was introduced to retain contrast-specific signals. Similarly, Colour TV better distinguishes tissue boundaries in lower-SNR images when there is clear delineation of tissues in a higher-SNR image. Any possible detrimental effects, when all images except one have noisy patterns in smoothly varying regions or across tissue boundaries, were prevented by the individual TV as it serves to reduce noise in individual images.

In practice, individual reconstruction of each acquisition in a multi-contrast protocol is better suited to online processing as it improves workflow by recovering the images for the a given contrast while data are being acquired for the next contrast. However, this does not preclude a workflow in which a given contrast is reconstructed without latency as the data become available to guide the prescription of later acquisitions in the protocol. At the end of the protocol, all acquisitions can still be jointly reconstructed for maximal image quality. This workflow would be similar to the one in Ref. [49] with the difference being that in SIMIT both images are jointly reconstructed instead of using an initially reconstructed image to improve the reconstruction of later images, without updating the first.

Selection of regularization parameters has a critical effect on the convergence behaviour and resultant image quality of regularized reconstructions. Each of the four regularization parameters used in SIMIT were separately varied until the average SSIM was reduced from 98% to below 95%. No significant variations in image quality were observed when the parameters were scaled up or down by an order of magnitude, suggesting that SIMIT is reasonably robust against variations in reconstruction parameters. The most sensitive parameter was that of group sparsity while the other parameters had broader margins. The optimal parameters may show larger deviations for body parts with substantially different tissue structure (e.g., the abdomen versus the brain). In such cases, parameters can be optimized a priori on a training dataset based on the anatomy of interest, yielding anatomy-specific sets of parameters.

The relative scales of image intensities and regularization parameters can affect the progression of iterative reconstructions. In case of a large mismatch in scale, it was observed that the updates in each iteration were either excessively small or large in magnitude, which caused all methods tested here to result in poor reconstructions. Therefore, intensity normalization was used to improve image quality and to ensure that similar ranges of regularization parameters work well across datasets. This is particularly important for joint reconstruction of multiple contrasts since image scales may vary significantly across acquisitions, and acquisitions with higher image intensities can dominate calculations of joint regularization terms such as joint sparsity or colour TV. To prevent potential scale-related biases, k-space data for each acquisition were normalized in this study, such that the respective fully-sampled reconstructed images are in the same range. Assuming similar initial noise levels, this normalization scales the noise-level for images with relatively low intensity upward, compared with the noise-floor of the images with higher intensity. To compensate for this increase in the noise level, we had to adjust the individual regularization terms for the T1-weighted image for all methods that use individual regularization to improve image quality. Even prior to adjustment, we prefer imbalanced noise levels across acquisitions to poor reconstruction quality.

SIMIT was demonstrated with both 1D and 2D undersampling, and thus it can be applied to both 2D and 3D imaging. Here we applied the regularization terms on cross-sectional images across the phase-encode directions. Alternatively, an entirely 3D optimization problem can be cast with regularization terms also incorporating tissue information along the readout dimension. In that case, group sparsity terms can be enforced across multiple cross-sections to further improve reconstruction performance.

## ACKNOWLEDGEMENTS

This work was supported in part by Turkish Scientific and Technological Research Council (TÜBİTAK) grant with Project #3151068, by a European Molecular Biology Organization Installation Grant (IG 3028), by TUBA GEBIP 2015 fellowships, and by BAGEP awards by the Science Academy.

## SUPPORTING INFORMATION

## THEORY

Here, we present a general formulation of the ADMM problem with multiple constraints and introduce the specific ADMM implementation for multi-contrast MRI.

### Generalized ADMM Formulation

To solve the optimization problem cast in Equations (1)-(2) in the main text, we devised an ADMM-based algorithm. In general, ADMM can be used to solve problems of type:

$$\min_{\boldsymbol{x},\boldsymbol{z}} f(\boldsymbol{x}) + g(\boldsymbol{z}) \qquad \text{[S1]}$$

$$subject\ to\ \boldsymbol{Px} + \boldsymbol{Qz} = \boldsymbol{c},$$

where an unconstrained multi-objective convex optimization problem is split via two variables $\boldsymbol{x}$ and $\boldsymbol{z}$, and a constraint is introduced with variables $\boldsymbol{P}, \boldsymbol{Q}$, and $\boldsymbol{c}$ that define the relationship between $\boldsymbol{x}$ and $\boldsymbol{z}$. Here we first show that the proposed SIMIT reconstruction can be cast in the form of Eq. [S1]. Note that the constrained optimization problem in Eq. (1) for multi-contrast images $(\boldsymbol{x})$ can be expressed as:

$$\min_{x} \xi_1\phi_1(\boldsymbol{x}) + \xi_2\phi_2(\boldsymbol{x}) + \cdots + \xi_m\phi_m(\boldsymbol{x})$$
$$subject\ to\ \left\|\boldsymbol{A}^{(i)}\boldsymbol{x}^{(i)} - \boldsymbol{y}^{(i)}\right\|_2 \leq \epsilon_i, i \in 1, \ldots, k, \qquad \text{[S2]}$$

where $k$ is the number of contrasts, $m$ is the number of regularization terms, $\phi_j$ denotes the j$^{\text{th}}$ regularization term (CTV, $\ell_{2,1}$, TV, or $\ell_1$) and $\xi_j$ is the corresponding regularization parameter, i.e. $\alpha_{CTV}$, $\gamma_{iTV}$, $\beta_{gL1}$, $\theta_{iL1}$. $\boldsymbol{A}^{(i)} = \boldsymbol{M}^{(i)}\boldsymbol{F}$ denotes the undersampled system observation matrix with $\boldsymbol{M}^{(i)}$ denoting the undersampling mask and $\boldsymbol{F}$ denoting the Fourier transformation matrix. $y^{(i)}$ is the acquired data, and $\left\|\boldsymbol{A}^{(i)}\boldsymbol{x}^{(i)} - \boldsymbol{y}^{(i)}\right\|_2 \leq \epsilon_i$ denotes the data fidelity constraint for the $i^{\text{th}}$ contrast.

To efficiently solve the optimization problem in Eq. [S2] using ADMM, we define $\boldsymbol{z}$ as the concatenation of k vectors:

$$\boldsymbol{z} = \left[\boldsymbol{z}^{(0)^T} \ldots \boldsymbol{z}^{(k)^T}\right]^T, \qquad \text{[S3]}$$

where the vector for each contrast $\boldsymbol{z}^{(i)}$ is defined as the concatenation of m sub-vectors $\boldsymbol{z}^{(i,t)}$ for each regularization term:

$$\boldsymbol{z}^{(i)} = \left[\boldsymbol{z}^{(i,0)^T} \boldsymbol{z}^{(i,1)^T} \ldots \boldsymbol{z}^{(i,m)^T}\right]^T. \qquad \text{[S4]}$$

Based on the definitions in Eqs. [S3]-[S4], here we propose solving:

$$\min_{x} \xi_1\phi_1\left(\left\{\boldsymbol{z}^{(i,1)}\right\}_{i=1,\ldots,k}\right) + \xi_2\phi_2\left(\left\{\boldsymbol{z}^{(i,2)}\right\}_{i=1,\ldots,k}\right) + \cdots + \xi_m\phi_m\left(\left\{\boldsymbol{z}^{(i,m)}\right\}_{i=1,\ldots,k}\right)$$
$$\text{subject to} \begin{cases} \left\|\boldsymbol{z}^{(i,0)} - \boldsymbol{y}^{(i)}\right\|_2 \leq \epsilon_i, & i = 1, \ldots, k \\ \boldsymbol{z}^{(i,0)} = \boldsymbol{A}^{(i)}\boldsymbol{x}^{(i)}, & i = 1, \ldots, k \\ \boldsymbol{z}^{(i,t)} = \boldsymbol{x}^{(i)}, & i = 1, \ldots, k, t = 1, \ldots, m \end{cases}. \qquad \text{[S5]}$$

Multiple sets of constraints are provided in Eq. [S5]. The first set is used to enforce data fidelity (as given in Eq. [S2]) on the first sub-vector of each $\boldsymbol{z}^{(i)}$, i.e., $\boldsymbol{z}^{(i,0)}$. The second set defines $\boldsymbol{z}^{(i,0)}$ as $\boldsymbol{A}^{(i)}\boldsymbol{x}^{(i)}$**.** The third set defines the remaining sub-vectors as $\boldsymbol{x}^{(i)}$ to pass those separately onto each regularization term. Here, we define a regularization function associated with the data fidelity constraint to treat it as a regularization term rather than a constraint. This change of variables does not change the solved problem and is equivalent to Eq. [S2]. We define $I_{E(\epsilon_i,\boldsymbol{I},\boldsymbol{y}^{(i)})}(\boldsymbol{v})$ as the indicator function of the constraint $\left(\left\|\boldsymbol{v} - \boldsymbol{y}^{(i)}\right\|_2 \leq \epsilon_i\right)$:

$$I_{E(\epsilon_i,\boldsymbol{I},\boldsymbol{y}^{(i)})}(\boldsymbol{v}) = \begin{cases} 0, & \left\|\boldsymbol{v}-\boldsymbol{y}^{(i)}\right\|_2 \leq \epsilon_i \\ \infty, & \left\|\boldsymbol{v}-\boldsymbol{y}^{(i)}\right\|_2 > \epsilon_i \end{cases}. \quad \text{[S6]}$$

Eq. [S5] is equivalent to Eq. [S1], and it can be cast into the same form with the following change of variables:

$$\begin{aligned} \boldsymbol{x} &= \left[\boldsymbol{x}^{(1)^T} \dots \boldsymbol{x}^{(m)^T}\right]^T, \\ f(\boldsymbol{x}) &= 0, \\ \boldsymbol{P}^{(i)} &= -\left[\left(\boldsymbol{A}^{(i)}\right)^T \boldsymbol{I} \dots \boldsymbol{I}\right]^T, \\ \boldsymbol{Q} &= \boldsymbol{I}, \\ \boldsymbol{c} &= 0, \\ g(\boldsymbol{z}) &= \textstyle\sum_{t=1}^{m} \xi_t \phi_t\left(\left\{\boldsymbol{z}^{(i,t)}\right\}_{i=1\dots k}\right) + \sum_{i=1}^{k} I_{E(\epsilon_i,\boldsymbol{I},\boldsymbol{y}^{(i)})}\left(\boldsymbol{z}^{(i,0)}\right), \end{aligned} \quad \text{[S7]}$$

where $\boldsymbol{P}$ is defined as a block diagonal matrix with $\boldsymbol{P}^{(i)}$ as its diagonal elements, $\boldsymbol{Q}$ is an identity matrix, and $\boldsymbol{c}$ is a zero vector.

Having shown that the proposed optimization problem for multi-contrast MRI can be cast in the general ADMM formulation of Eq. [S1], we derive the following update rules:

$$\boldsymbol{x}_{n+1} = \underset{\boldsymbol{x}}{\operatorname{argmin}} \|\boldsymbol{z}_n - \boldsymbol{P}\boldsymbol{x} + \boldsymbol{d}_n\|_2^2 \quad \text{[S8]}$$

$$\boldsymbol{z}_{n+1} = \underset{\boldsymbol{z}}{\operatorname{argmin}} \left[\sum_{t=1}^{m} \xi_t \phi_t\left(\left\{\boldsymbol{z}^{(i,t)}\right\}_{i=1\dots k}\right) + \sum_{i=1}^{k} I_{E(\epsilon_i,\boldsymbol{I},\boldsymbol{y}^{(i)})}\left(\boldsymbol{z}^{(i,0)}\right) + \frac{\mu}{2}\|\boldsymbol{z} - \boldsymbol{P}\boldsymbol{x}_{n+1} + \boldsymbol{d}_n\|_2^2\right] \quad \text{[S9]}$$

$$\boldsymbol{d}_{n+1} = \boldsymbol{d}_n + \boldsymbol{z}_{n+1} - \boldsymbol{P}\boldsymbol{x}_{n+1} \quad \text{[S10]}$$

where the subscript $n$ denotes the state of any given variable at iteration $n$, and $\boldsymbol{d}$ denotes the dual variable associated with the Lagrangian of Eq. [S1]. The problem in Eq. [S6] is a simple least squares problem with block-diagonal entries. It can be separated and solved for each contrast as:

$$\boldsymbol{x}_{n+1}^{(i)} = \underset{\boldsymbol{x}^{(i)}}{\operatorname{argmin}} \left\|\boldsymbol{P}^{(i)}\boldsymbol{x}^{(i)} - \left(\boldsymbol{z}_n^{(i)} + \boldsymbol{d}_n^{(i)}\right)\right\|_2^2, \quad \text{[S11]}$$

$$= \left(\boldsymbol{P}^{(i)^H}\boldsymbol{P}^{(i)}\right)^{-1}\boldsymbol{P}^{(i)^H}\left(\boldsymbol{z}_n^{(i)} + \boldsymbol{d}_n^{(i)}\right), \quad \text{[S12]}$$

$$= \left(m\boldsymbol{I} + \boldsymbol{A}^{(i)^H}\boldsymbol{A}^{(i)}\right)^{-1}\left(\boldsymbol{A}^{(i)^H}\left(\boldsymbol{z}_n^{(i,0)} + \boldsymbol{d}_n^{(i,0)}\right) + \sum_{t=1}^{m}\left(\boldsymbol{z}_n^{(i,t)} + \boldsymbol{d}_n^{(i,t)}\right)\right). \quad \text{[S13]}$$

Hence the update equation for $\boldsymbol{x}$ can be decomposed into $k$ separate least-squares problems. For MRI data from a single receiver coil, $\boldsymbol{P}^{(i)}$'s are in the form of masked unitary transforms. In this case, each least squares operation to calculate $\boldsymbol{x}_{n+1}^{(i)}$ and $\boldsymbol{A}^{(i)}\boldsymbol{x}_{n+1}^{(i)}$ can be implemented using several simple element-wise operations (O(N)) and 2 FFT operations per contrast per iteration (O(NlogN)) [1]. The updates in Eq. [S9] can also be separated for each regularization term $\phi_t(\cdot)$ and the indicator $I_{E(\epsilon_i,\boldsymbol{I},\boldsymbol{y}^{(i)})}(\boldsymbol{z}^{(i,0)})$ . For the sub-vector associated with data fidelity constraint the update becomes:

$$\boldsymbol{z}_{n+1}^{(i,0)} = \underset{\boldsymbol{z}^{(i,0)}}{\operatorname{argmin}}\, I_{E(\epsilon_i,\boldsymbol{I},\boldsymbol{y}^{(i)})}(\boldsymbol{z}^{(i,0)}) + \frac{\mu}{2}\left\|\boldsymbol{z}^{(i,0)} - \boldsymbol{A}^{(i)}\boldsymbol{x}_{n+1}^{(i)} + \boldsymbol{d}_n^{(i,0)}\right\|_2^2, \qquad \text{[S14]}$$

$$\boldsymbol{d}_{n+1}^{(i,0)} = \boldsymbol{d}_n^{(i,0)} + \boldsymbol{z}_{n+1}^{(i,0)} - \boldsymbol{A}^{(i)}\boldsymbol{x}_{n+1}^{(i)}. \qquad \text{[S15]}$$

Eq. [S14] is a simple projection onto an $\ell_2$-norm hyper-sphere [2]. Here, the variable $\mu$ is the augmented Lagrangian parameter, and is associated with the inverse of the step size for the algorithm. For the remaining sub-vectors, the z-update step for each regularization terms in Eq. [S9] becomes:

$$\left\{\boldsymbol{z}_{n+1}^{(i,t)}\right\}_{i=1\ldots k} = \underset{z}{\operatorname{argmin}}\, \xi_t\phi_t\left(\left\{\boldsymbol{z}^{(i,t)}\right\}_{i=1\ldots k}\right) + \frac{\mu}{2}\left\|\left\{\boldsymbol{z}^{(i,t)} - \boldsymbol{x}_{n+1}^{(i)} + \boldsymbol{d}_n^{(i,t)}\right\}_{i=1\ldots k}\right\|_2^2, \qquad \text{[S16]}$$

$$\boldsymbol{d}_{n+1}^{(i,t)} = \boldsymbol{d}_n^{(i,t)} + \boldsymbol{z}_{n+1}^{(i,t)} - \boldsymbol{x}_{n+1}^{(i)}. \qquad \text{[S17]}$$

The operation in Eq. [S16] is called the Moreau proximal mapping function [3], the result of which we shall denote $\Psi_{\frac{\xi_t\phi_t}{\mu}}\left(\left\{\boldsymbol{x}_{n+1}^{(i)} - \boldsymbol{d}_n^{(i,t)}\right\}_{i=1\ldots k}\right)$.

ADMM is known to converge under mild conditions [3]. For non-convex problems, if the exact solution of each sub-problem is known, then the algorithm converges to a local minimum. Step size parameter $1/\mu$ determines the rate of convergence; a smaller $\mu$ means larger steps and faster convergence. However, the algorithm may diverge for very small $\mu$. Therefore, the step size should be carefully selected to ensure good convergence behaviour [3]. An automated way of selecting this parameter is given in [4].

**Extension to Parallel Imaging + Compressed Sensing**

To extend the described algorithm to parallel imaging, coil sensitivities need to be incorporated. This was achieved by representing the encoding matrix $\boldsymbol{A}^{(i)}$ as a concatenation of the encoding matrix for each coil as $\boldsymbol{A}^{(i)} = \begin{bmatrix}\boldsymbol{A}^{(i,1)}\\ \cdots \\ \boldsymbol{A}^{(i,N_c)}\end{bmatrix} = \begin{bmatrix}\boldsymbol{M}^{(i)}\boldsymbol{F}\boldsymbol{C}^{(1)}\\ \cdots \\ \boldsymbol{M}^{(i)}\boldsymbol{F}\boldsymbol{C}^{(N_c)}\end{bmatrix}$, where $\boldsymbol{C}^{(j)}$ is the coil sensitivity for channel $j$, and separating each dual variable $\boldsymbol{z}^{(i,0)}$ into $N_c$ parts as $\boldsymbol{z}^{(i,0)} = \left[\boldsymbol{z}^{(i,0,1)^T} \ldots \boldsymbol{z}^{(i,0,N_c)^T}\right]^T$, where $N_c$ is the number of coils. Then, the matrix associated with ADMM in Eq. [S7] and the indicator function in Eq. [S6] become:

$$\boldsymbol{P}^{(i)} = -\left[\left(\boldsymbol{FC}^{(1)} \cdots\ \boldsymbol{FC}^{(l)}\right)^T \boldsymbol{I} \ldots \boldsymbol{I}\right]^T, \quad \text{[S18]}$$

$$I_{E(\epsilon_{ij},\boldsymbol{I},\boldsymbol{y}^{(i)})}\left(\boldsymbol{z}^{(i,0)}\right) = \begin{cases} 0, & \left\|\boldsymbol{M}^{(i)}\boldsymbol{z}^{(i,0,j)} - \boldsymbol{y}^{(i,j)}\right\|_2 < \epsilon_{i,j} \quad \forall j \\ \infty, & \text{otherwise.} \end{cases} \quad \text{[S19]}$$

This requires two updates to given equations. First, Eq. [S11] should be updated to reflect this change in $\boldsymbol{P}^{(i)}$.

$$\boldsymbol{x}_{n+1}^{(i)} = \left(\boldsymbol{P}^{(i)^H}\boldsymbol{P}^{(i)}\right)^{-1} \boldsymbol{P}^{(i)^H}\left(\boldsymbol{z}_n^{(i)} + \boldsymbol{d}_n^{(i)}\right), \quad \text{[S20]}$$

$$= \left(m\boldsymbol{I} + \sum_{\boldsymbol{j}=\boldsymbol{1}}^{N_c}\left|\boldsymbol{C}^{(j)}\right|^{\boldsymbol{2}}\right)^{-1} \left(\sum_{j=1}^{N_c}\boldsymbol{C}^{(j)^H}\left(\boldsymbol{z}_n^{(i,0,j)} + \boldsymbol{d}_n^{(i,0,j)}\right) + \sum_{t=1}^{m}\left(\boldsymbol{z}_n^{(i,t)} + \boldsymbol{d}_n^{(i,t)}\right)\right), \quad \text{[S21]}$$

All operations given in Eq. [S21] are with diagonal matrices, and can be implemented using simple element-wise operations. Next, by extending the data fidelity to handle each coil separately using these definitions, the block diagonal structure in x-update step of Eq. [S12] is preserved, and this step can still be implemented in O(NlogN) operations using only simple FFT operations. Moreover, the projection step is still the same except now some elements of $\boldsymbol{z}$ is kept constant. Final necessary changes are, Eq. [S16] becomes:

$$\boldsymbol{z}_{n+1}^{(i,0)} = \underset{\boldsymbol{z}^{(i,0)}}{\operatorname{argmin}}\ I_{E(\epsilon_i,\boldsymbol{I},\boldsymbol{y}^{(i)})}\left(\boldsymbol{z}^{(i,0)}\right) + \frac{\mu}{2}\left\|\boldsymbol{z}^{(i,0)} - \begin{bmatrix}\boldsymbol{FC}^{(1)} \\ \cdots \\ \boldsymbol{FC}^{(N_c)}\end{bmatrix}\boldsymbol{x}_{n+1}^{(i)} + \boldsymbol{d}_n^{(i,0)}\right\|_2^2, \quad \text{[S22]}$$

and Eq. [S15] becomes:

$$\boldsymbol{d}_{n+1}^{(i,0)} = \boldsymbol{d}_n^{(i,0)} + \boldsymbol{z}_{n+1}^{(i,0)} - \begin{bmatrix}\boldsymbol{FC}^{(1)} \\ \cdots \\ \boldsymbol{FC}^{(N_c)}\end{bmatrix}\boldsymbol{x}_{n+1}^{(i)}. \quad \text{[S23]}$$

**Solving SIMIT using ADMM**

To efficiently solve the SIMIT problem using the adapted ADMM algorithm described above, proximal mapping functions are needed that yield the solution of each subproblem associated with each regularization term. The proximal mapping function of the individual $\ell_1$-norm is a simple element-wise operator known as soft-thresholding [1]:

$$\Psi_{\frac{\xi_t\|\cdot\|_1}{\mu}}(\boldsymbol{v}) = \exp\{1j\angle\boldsymbol{v}\} \cdot \max\left\{0, |\boldsymbol{v}| - \frac{\xi_t}{\mu}\right\}. \quad \text{[S24]}$$

The proximal mapping function of scaled group sparsity can be derived as:

$$\Psi_{\frac{\xi_t\|\cdot\|_{2,1}}{\mu}}(\boldsymbol{v}) = \left\{\boldsymbol{v}^{(i)} \cdot \max\left\{0, 1 - \frac{\xi_t}{\mu\|\boldsymbol{v}\|_2}\right\}\right\}_{i=1\cdots k}, \quad \text{[S25]}$$

where $||\boldsymbol{v}||_2$ is defined across the contrasts. Note that, both definitions retain the phase of the input-function, and therefore, are readily applicable to complex images.

For TV and CTV functions, proximal mapping functions rely on an algorithm that minimizes the dual function as proposed by Chambolle [5] for individual TV, and Bresson for CTV [6]. This study uses TV and CTV regularization terms on the magnitude of the image. For both algorithms proximal mapping functions associated with real-valued inputs with magnitude of the input vector were used while the phase was retained separately, similar to Eq. [S24] [Proximal mapping functions for TV and CTV are not derived here, since those were rigorously derived in Ref. [1,5,6]].

Finally, the regularization weights $\alpha_{CTV}$, $\gamma_{iTV}$, $\beta_{gL1}$, $\theta_{iL1}$ are assigned to the $\xi_i$'s that correspond to their respective regularization terms.

**Algorithm**

Set iteration variable $n$=0, choose step size $\mu$ > 0

Initialize the dual variables $\boldsymbol{z}_{\mathbf{0}}^{(i,t)}, \boldsymbol{d}_{\mathbf{0}}^{(i,t)}$

**Repeat**

**for** *i* = *1…k* where *k* is the number of contrasts

Update image $\boldsymbol{x}_{n+1}^{(i)}$ via Eq. [S13] (single-channel) or Eq. [S21] (multi-channel)

Update $\boldsymbol{z}_{n+1}^{(i,0)}$ via Eq. [S14] (single-channel) or Eq. [S22] (multi-channel)

Update $\boldsymbol{d}_{n+1}^{(i,0)}$ via Eq. [S15] (single-channel) or Eq. [S23] (multi-channel)

**end for**

**for** *t* = *1…m* where *m* is the number of regularization functions

Update $\left\{\boldsymbol{z}_{n+1}^{(i,t)}\right\}_{i=1\ldots k}$ via Eq. [S16] for each contrast *i* = *1…k*

Update $\boldsymbol{d}_{n+1}^{(i,t)}$ via Eq. [S17], for each contrast *i* = *1…k*

**end for**

Increment iteration number n ← n+1

**Until** some stopping criterion is satisfied.

At iteration n+1, $k(N_c + m)$ instances of the variables $\boldsymbol{z}_{n+1}^{(.)}$ and $\boldsymbol{d}_{n+1}^{(.)}$ are created, one for data fidelity on each channel per contrast (and one for each regularization term, per contrast. The algorithm finds the $\boldsymbol{z}_{(.)}^{(.)}$ that minimize the functions in Eq. [S14] (or Eq. [S20] for multi-channel) and Eq. [S16] using the current contrast images ($\boldsymbol{x}_{n+1}^{(i)}$) and the previous instance of $\boldsymbol{d}_{n}^{(.)}$; update $\boldsymbol{d}_{n+1}^{(.)}$ based on the current image and current $\boldsymbol{z}_{n+1}^{(.)}$, and then combine all the new instances of $\boldsymbol{z}_{n+1}^{(.)}$ and $\boldsymbol{d}_{n+1}^{(.)}$ to update the contrast images.

## METHODS

The proposed joint reconstruction method (SIMIT) was compared to seven CS methods for reference including Sparse MRI [7], TVCMRI [8], RecPF [9], GSMRI [10], FCSA [11] and FCSA-MT [12], and a modified version that only included individual regularization terms (Indiv-only) [1] for a single-channel receiver coil. To compare the methods, the numerical phantom and in-vivo data from one participant were used (details of the datasets are given in the main text). Parameters for all methods were optimized using the procedure outlined in the main text. The automatically optimized parameters yielded inferior SSIM values for FCSA and FCSA-MT, which did not improve through manual optimization of the parameters. Therefore, the parameters given in Ref. [12] were used.

| Optimized parameters | $\alpha_{CTV}$ | $\beta_{GL1}$ | $\gamma_{iTV}$ | $\theta_{iL1}$ |
|---|---|---|---|---|
| SparseMRI | | | 0.012 | 0.01 |
| TVCMRI | | | 0.355 | 0.696 |
| RecPF | | | 0.419 | 0.04 |
| GSMRI | | 0.035 | | |
| FCSA | | | 0.01 | 0.035 |
| FCSA-MT | 0.01 | 0.035 | | |
| Indiv-only | | | 0.021 | 1.142 |
| SIMIT | $0.19/\sqrt{k}$ | $0.51/\sqrt{k}$ | 0.11/k | 9.13/k |

***Table SI-1:*** *Optimized weight parameters for constraint functions for all methods. For the proposed method, $k$ denotes the number of contrasts to be jointly reconstructed.*

### Numerical Phantom

The methods were compared in terms of SSIM and pSNR. To assess the stability of reconstruction performance across undersampling masks and noise distributions, a Monte-Carlo simulation with 250 runs was performed with independent instances of undersampling masks and noise. Runtimes (excluding data-preparation) at each iteration were measured with the cputime command in Matlab (which excludes any parallel computing capabilities) and averaged across runs. Image quality metrics averaged across runs were plotted as a function of cumulative runtime for each method. These comparisons were made using only three contrasts (PD-, T1- and T2-weighted images).

To assess the performance of the methods for different acceleration rates, all methods were compared in terms of nRMSE for acceleration rates between R=2 and R=5 for 1D-acceleration and between R=2 and R=15 for 2D-acceleration using the five-contrast dataset.

### In-vivo Data

To compare with the reference methods [1,7-12] that were developed for a single-channel receiver coil, reconstructions were performed on acquisitions of retrospectively 2D-undersampled (R=3) three-contrast data of one of the participants. Reconstructions were performed separately for each channel of the 32-channel receiver-array and combined afterwards [13].

## RESULTS

### Single channel comparisons - Numerical Phantom

Figure SI-2 shows the image quality metrics pSNR and SSIM of multi-contrast reconstructions as a function of reconstruction time. Note that Indiv-only is omitted here as SIMIT and Indiv-only are already compared in the main text. The proposed method, SIMIT, achieves superior image quality than all alternative reconstructions upon convergence (Figure SI-2a). Note that FCSA and FCSA-MT yield substantially lower quality compared to remaining reconstructions although a broad range of regularization parameters were considered. The closest competitor to SIMIT is RecPF. While SIMIT has $32.5 \pm 0.1$ dB (mean $\pm$ standard deviation) pSNR and $97.5 \pm 0.1\%$ SSIM, RecPF had $30.4 \pm 0.2$ dB pSNR and $95.5 \pm 0.3\%$ SSIM. RecPF rapidly converges to a stable solution, but SIMIT surpasses RecPF in terms of image quality within 11 seconds of runtime. The superior performance of SIMIT is robust against variability in undersampling masks and noise instances (Figure SI-2b).

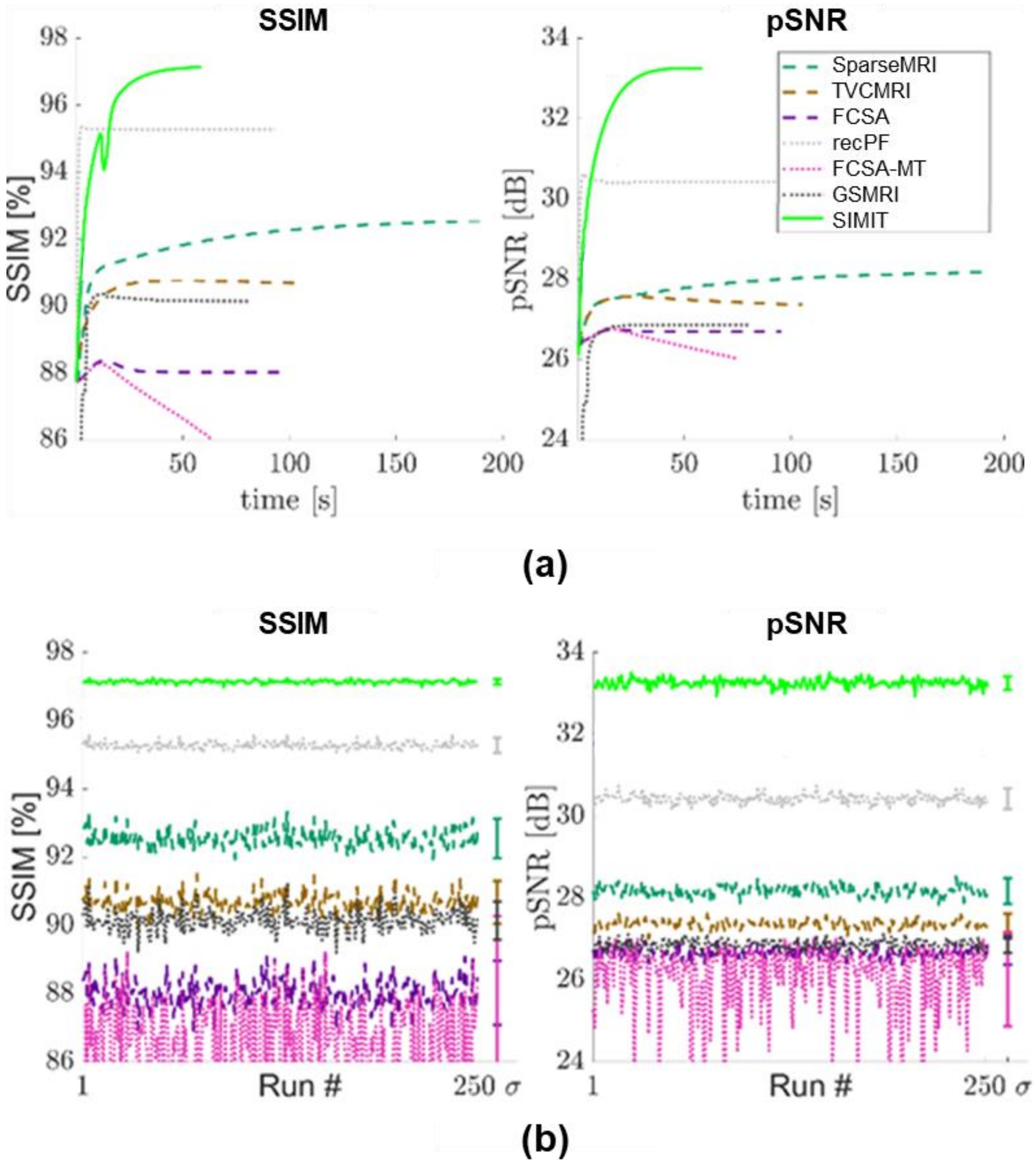

*__Figure SI-2:__ Image metrics for the Monte-Carlo runs on the numerical phantom for R=3. In each run, a different undersampling mask and noise instance was used. Masks and noise instances were kept identical across methods*

*within runs.* ***(a)*** *Image metrics averaged over the runs are given as a function of reconstruction time.* ***(b)*** *Image metrics for the final images are given with respect to the Monte-Carlo runs. Axes were adjusted to show a smaller range of values to improve comparison among high performing methods. The proposed method SIMIT improves all metrics compared to alternative reconstructions. The standard deviations of each metric plotted on the right in each figure in panel (b) show that SIMIT is more robust against variations in the undersampling masks and noise distributions, yielding a more stable performance.*

Reconstructed images are given in Figures SI-3 and SI-4. Difference images between the reconstructed and the fully-sampled reference images show that SIMIT has lower error compared to all other reconstructions. Note that, 1D-acceleration in the AP direction leads to residual aliasing artefacts with all reference methods, particularly in PD-weighted images. In contrast, SIMIT successfully suppresses residual artefacts to achieve improved tissue depiction.

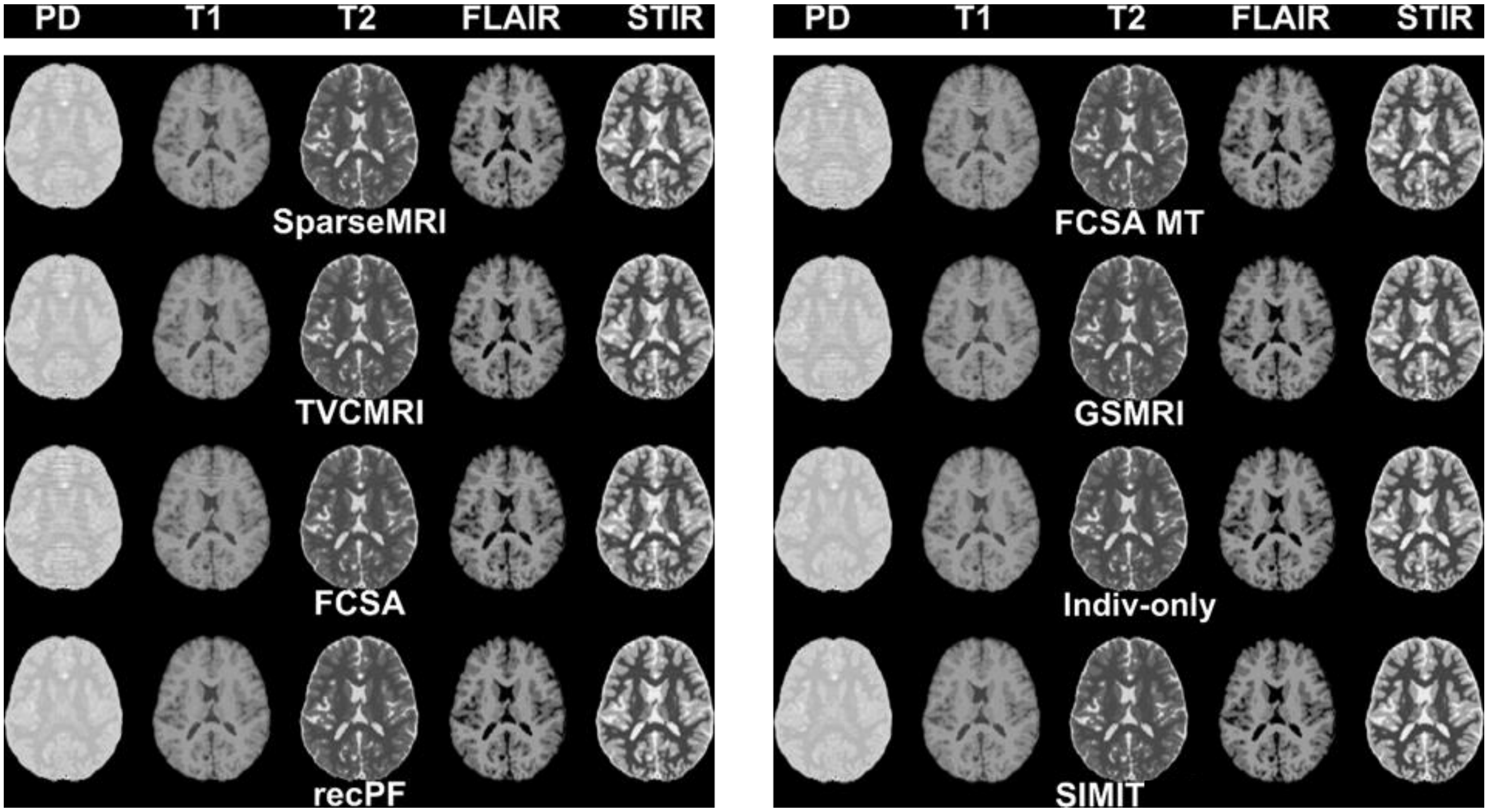

***Figure SI-3:*** *Multi-contrast reconstructions for the numerical phantom with 1D-acceleration in the AP direction and R = 3. Contrasts shown are PD, T1, T2-weighted, FLAIR and STIR. SIMIT visibly improves image quality compared to alternative reconstructions.*

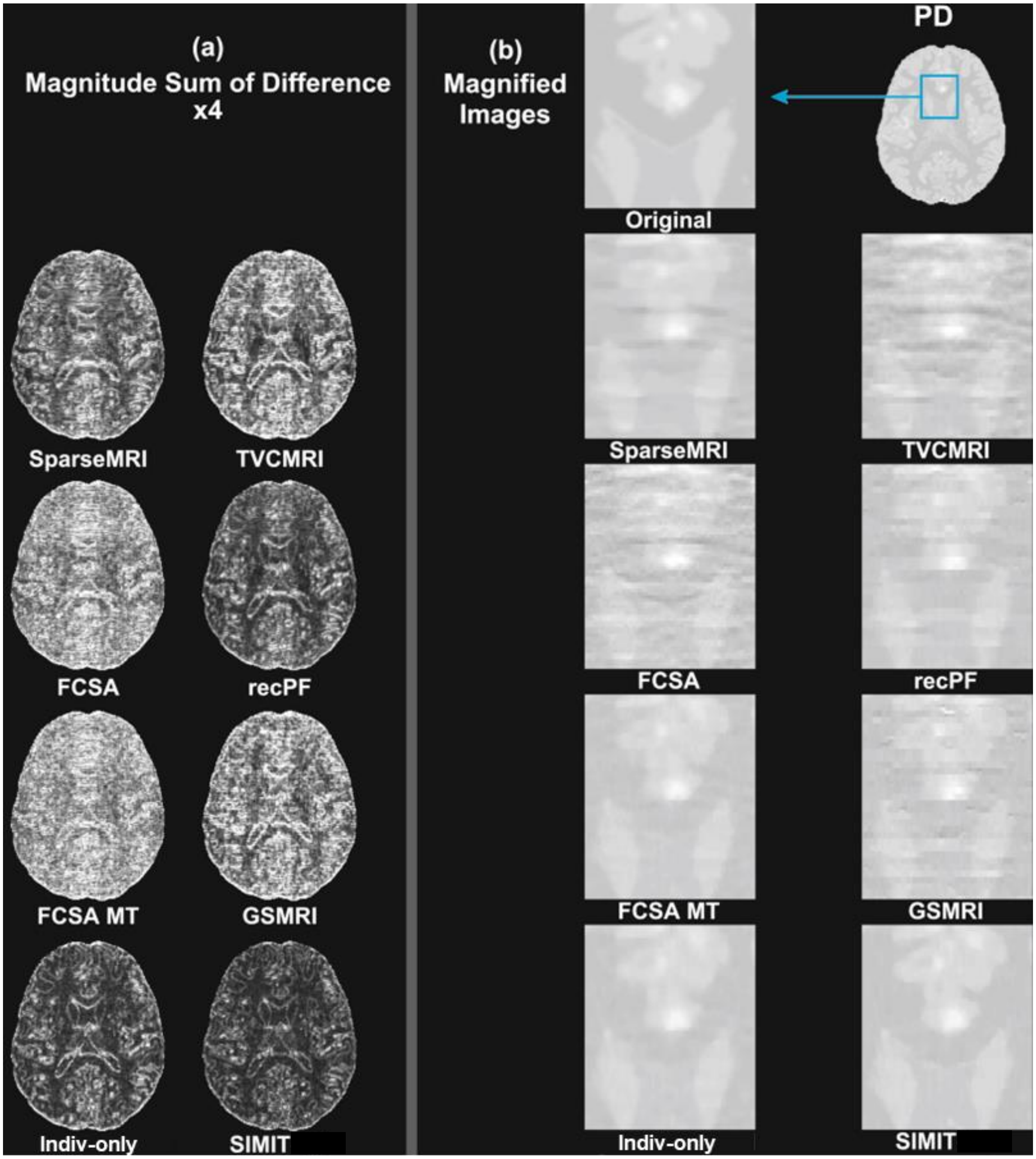

***Figure SI-4:*** *Multi-contrast reconstructions for the numerical phantom with 1D-acceleration in the AP direction and R = 3.* ***(a)*** *Differences between the reconstructed images and the fully-sampled reference images were calculated for each contrast. Magnitude difference images were summed across contrasts and 4x intensified, and then, displayed in the same scale as in Figure SI-3b.* ***(b)*** *Magnified PD-weighted images show a region of interest in the upper half of the FOV. Compared to other methods, SIMIT offers superior suppression of residual artefacts.*

Figure SI-5 compares the single-channel reconstruction methods for various 1D- and 2D-acceleration rates and shows that SIMIT consistently yields lower image nRMSE, for each image and on the average.

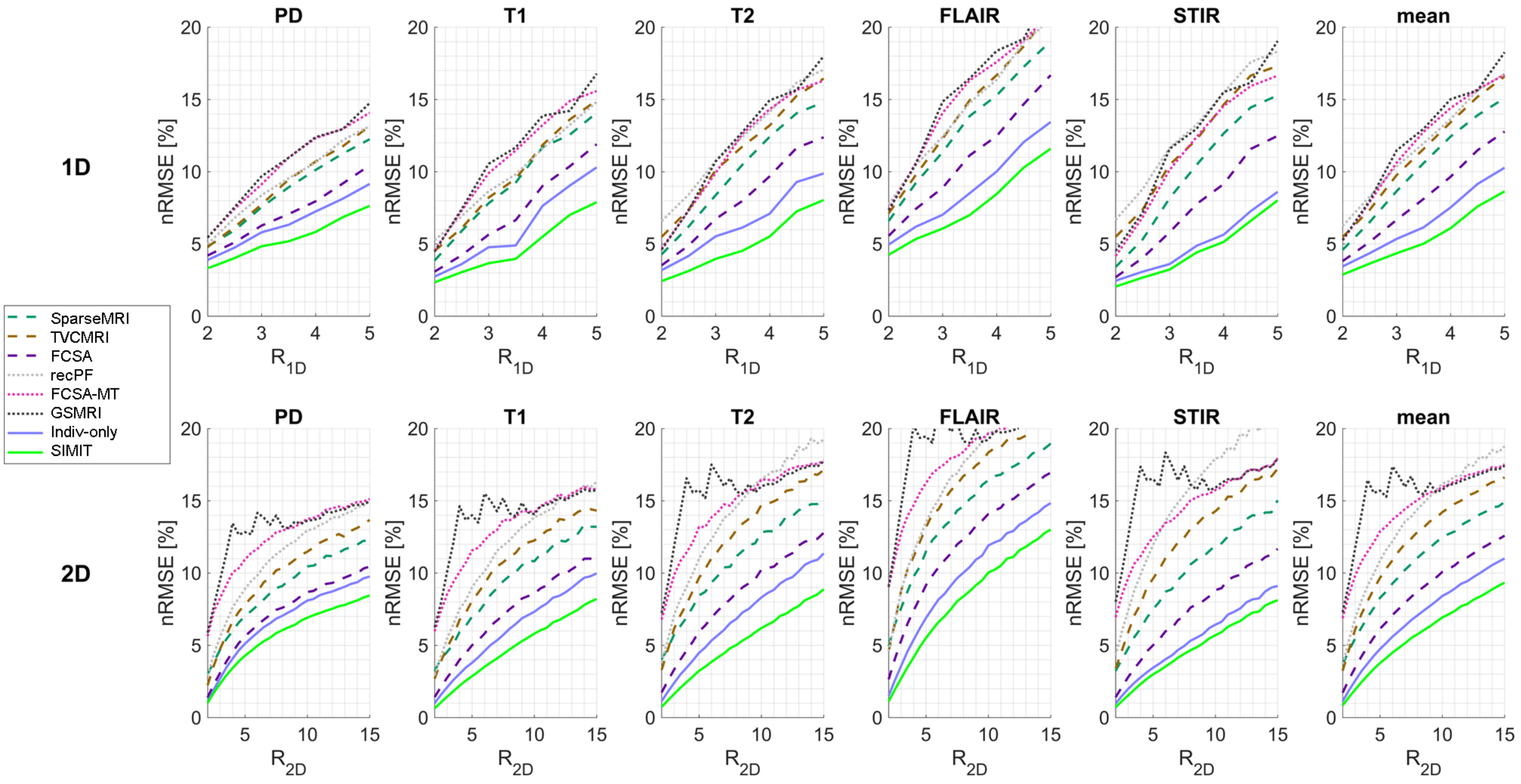

***Figure SI-5:*** *Methods are compared in terms of nRMSE for various acceleration rates, ranging between R=2 and R=5 for one-dimensional and R=2 and R=15 for two-dimensional acceleration, for a five-contrast imaging case with PD, T1-weighted, T2-weighted, FLAIR and STIR images. Undersampling masks were varied across contrasts, but same set of masks were used for each method. Image error (nRMSE) is shown separately for each contrast and as a mean across contrasts. SIMIT consistently provided reconstructions with lower image error for all contrasts and acceleration rates.*

**Single channel comparisons – In-vivo Data**

PD-, T1-, and T2-weighted *in-vivo* acquisitions from a single subject were reconstructed via SIMIT and reference methods. Representative reconstructions for 2D acceleration with R=3 are shown in Figures SI-6 and SI-7. SIMIT yields more detailed depiction of tissue structure compared to both individual and joint reconstruction methods (Figure SI-7). This is reflected in the SSIM values: SIMIT yields 95.8% SSIM, 33.9 dB pSNR while RecPF has 90.1% SSIM, 33.4 dB pSNR and TVCMRI has 87.8% SSIM, 31.8 dB pSNR.

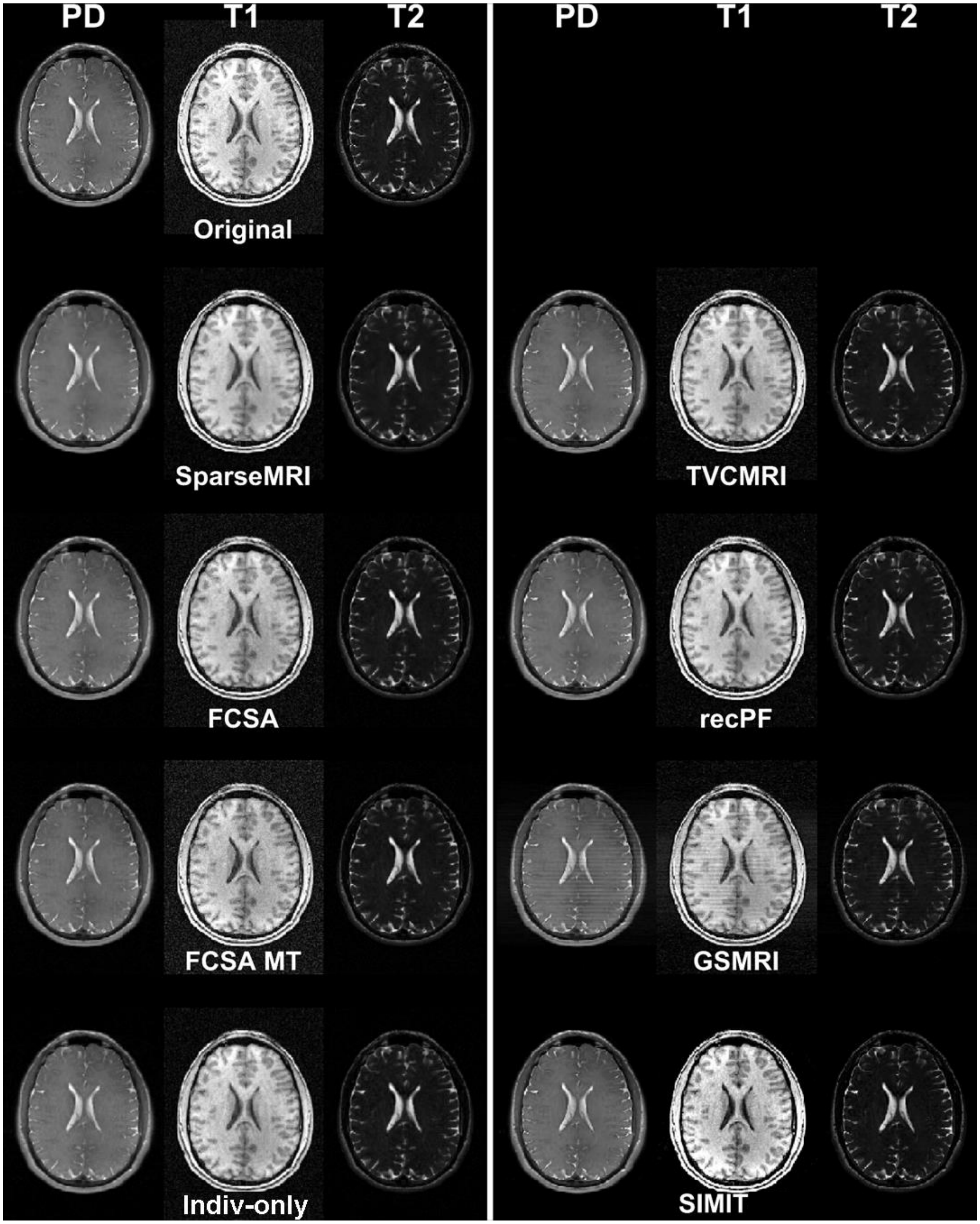

***Figure SI-6:*** *Reconstructed images for the in-vivo data with 2D-acceleration and R = 3. The fully-sampled reference images are shown along with SIMIT and seven other state-of-the-art reconstruction methods. SparseMRI, TVCMRI, FCSA, RecPF, and Indiv-only yield reconstructions with apparent losses in image sharpness, especially in T1-weighted images. GSMRI yields strong striping artefacts, while FCSA-MT shows a high-level of residual noise. In contrast, SIMIT achieves improved tissue delineation with relatively limited noise amplification compared to FCSA-MT.*

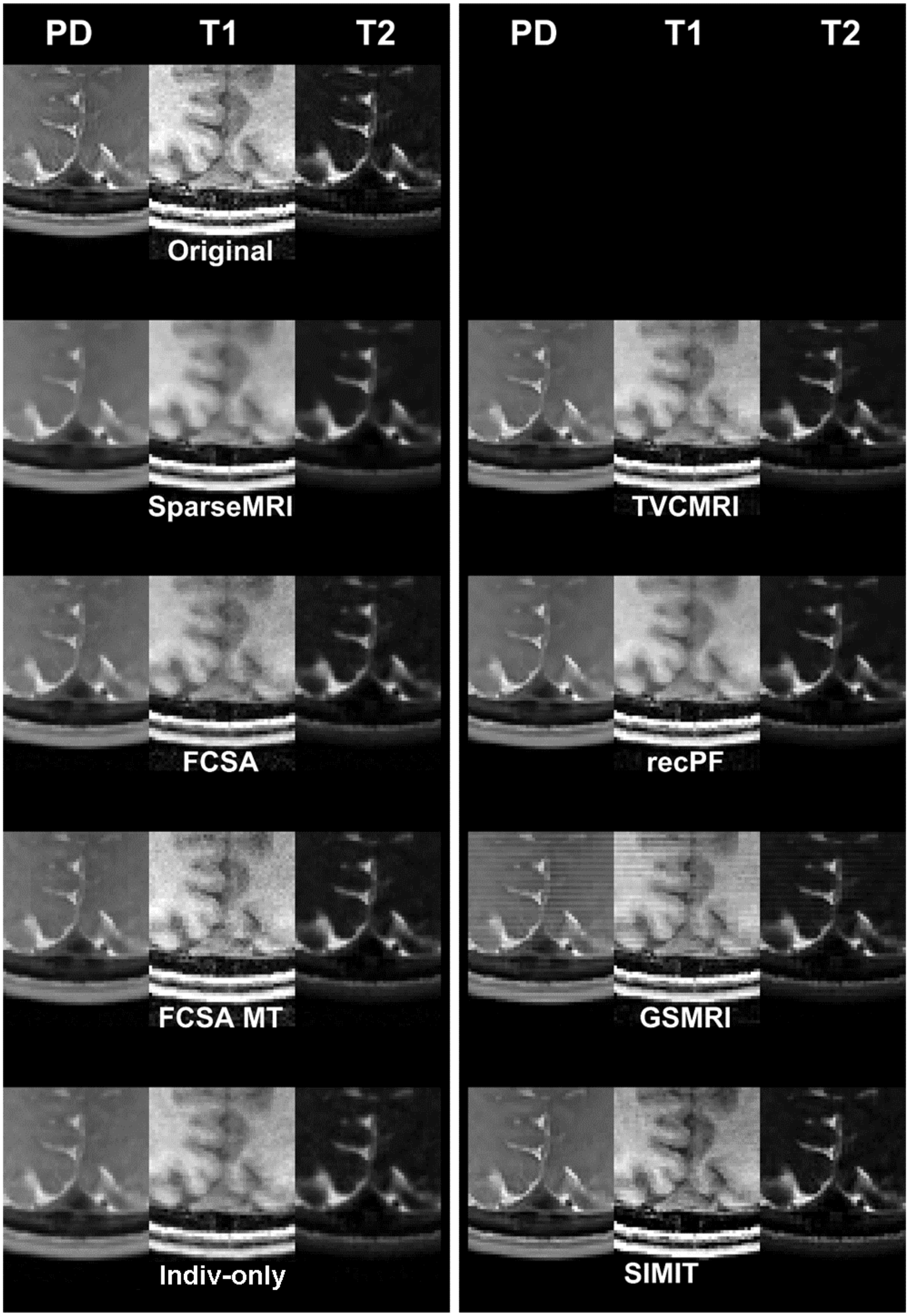

***Figure SI-7:*** *Reconstructed images for the in-vivo data with 2D-acceleration and R = 3. Magnified images from a region of interest in the posterior part of the FOV are given. SIMIT visually improves tissue delineation and image sharpness compared to all methods including FCSA-MT. SIMIT also improves image sharpness compared to its individual implementation, Indiv-only.*

**Multi-channel comparisons**

Contrasts have different pSNR and SSIM levels for the same acceleration rates in Figure 4 (main text) as the reconstruction error depends on the image content. While the tissue boundaries and therefore the underlying frequency content are the same for all contrasts, the energy at each spatial frequency is different. SSIM is a quantitative metric that captures the perceptual quality of reconstructed images, and is therefore more sensitive to local texture information compared to MSE-based metrics such as pSNR. FLAIR images tend to have sharper transitions across tissue boundaries and elevated high-spatial-frequency content compared to other contrast examined (e.g., PD images). Since reconstruction performance is expectedly poorest for high-spatial-frequency samples that are heavily undersampled in accelerated MRI, it is reasonable that FLAIR images have lower SSIM values for the same acceleration factor, leading to the differences in pSNR and SSIM levels (Figure SI-8).

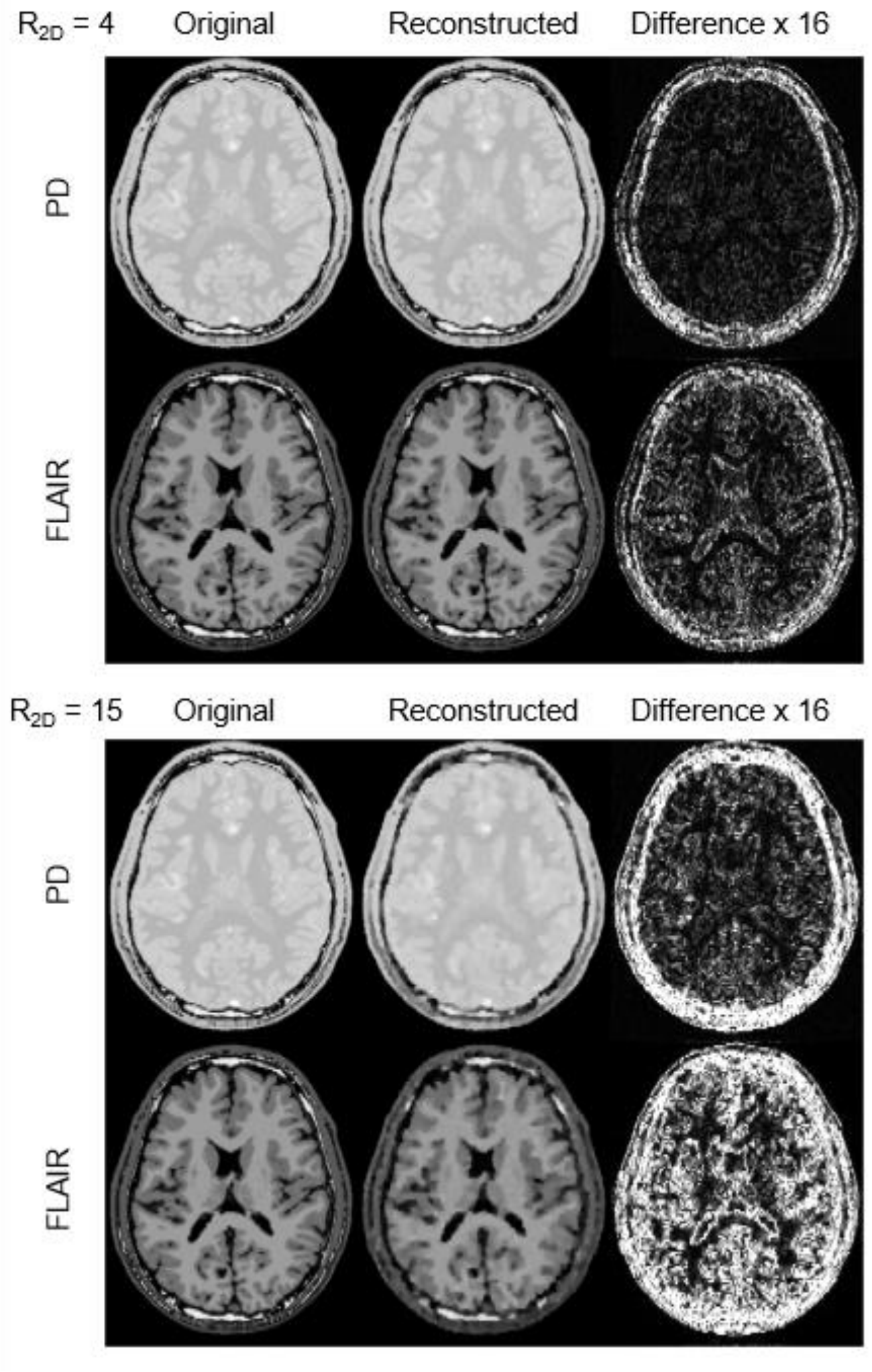

***Figure SI-8:*** *The reconstruction performance depends on the image content. Because FLAIR-images have higher intensity high-frequency variations inside the brain than PD-images, the reconstruction error inside the brain at the same acceleration factors is higher, which leads to lower performance in terms of pSNR and SSIM in Figure 4.*

## REFERENCES FOR SUPPORTING INFORMATION

## LIST OF ABBREVIATIONS

MRI: magnetic resonance imaging
PI: parallel imaging
CS: compressive sensing
SAR: specific absorption rate
TV: Total Variation
SIMIT: SIMultaneous use of Individual and joinT regularization terms for joint CS-PI reconstruction
ADMM: Alternating Direction Method of Multipliers
Indiv-only: Reconstruction method that uses only individual $\ell_1$-sparsity and TV terms
FCSA-MT: Multi Contrast FCSA
CTV: colour TV
gL1: Group $\ell_1$-sparsity, implemented as an $\ell_{2,1}$-norm
iTV: Individual TV
iL1: Individual $\ell_1$-sparsity
pSNR: peak signal-to-noise ratio
PD / T1 / T2: Proton density / T1- / T2- weighted
1D / 2D: one- / two- dimensional
Joint-only: Reconstruction method that uses only joint terms CTV and group $\ell_1$-sparsity
ESPIRiT: Eigenvalue-based implementation of Iterative self-consistent parallel imaging reconstruction from arbitrary k-space
TVCMRI: Total Variation $\ell_1$ Compressed MR Imaging
RecPF: reconstruction from partial Fourier data (RecPF)
GSMRI: Group-Sparse MRI
FCSA: Fast Composite Splitting Algorithm

TE / TI / TR: echo / inversion / repetition time
FLAIR: fluid-attenuated inversion recovery
STIR: short-time inversion recovery
FOV: field-of-view
SSIM: structural similarity
R: Acceleration rate
ROI: Region-of-interest